\documentclass[aps,prd,preprint,tightenlines,superscriptaddress,
amsfonts,amssymb,amsmath,byrevtex,showpacs,nofootinbib]{revtex4-1}
\usepackage[polish,english]{babel}
\usepackage[applemac]{inputenc}
\usepackage[T1]{fontenc}
\usepackage{latexsym}
\usepackage{graphicx}
\usepackage{geometry}
\usepackage{epstopdf}
\usepackage{subfig}
\usepackage{color}

\usepackage{xcolor}
\usepackage{pdfsync}
\usepackage{fancyhdr}
\usepackage{makeidx}
\usepackage[breaklinks,colorlinks,backref]{hyperref}
\hypersetup{
    colorlinks, 
    linktoc=all, 
    linkcolor=red,  
  citecolor=hyptxt,
  urlcolor=blue}
  \hypersetup{
  citebordercolor=,
  filebordercolor=green,
  linkbordercolor=blue
}
\definecolor{hyptxt}{rgb}{0.7, 0.4, 0.9}
\usepackage{bibtopic}
\newcommand{\bea}{\begin{eqnarray}}
\newcommand{\eea}{\end{eqnarray}}

\newcommand{\dR}{\mathbb R}
\newcommand{\dC}{\mathbb C}

\newcommand{\id}{\mathbb I}

\newcommand{\be}{\begin{equation}}
\newcommand{\ee}{\end{equation}}

\newcommand{\I}{\mathbb I}

\newcommand{\ket}[1]{|\kern.3ex#1\kern.3ex\rangle}
\newcommand{\bra}[1]{\langle\kern.3ex #1 \kern.3ex|}
\newcommand{\scalar}[2]{\langle\kern.3ex #1 \kern.3ex|\kern.3ex#2\kern.3ex\rangle}
\newcommand{\norm}[1]{\|\kern.3ex#1\kern.3ex \|}

\newcommand{\UnitOp}{\hat{\mathbb{I}}} 
\newcommand{\Group}[1]{\mathrm{#1}} 
\newcommand{\Bra}[1]{\langle #1 \vert} 
\newcommand{\Ket}[1]{\vert #1 \rangle} 
\newcommand{\BraKet}[2]{\langle #1 \vert #2 \rangle} 

\begin{document}

\title{Quantum Belinski-Khalatnikov-Lifshitz scenario}

\author{Andrzej G\'{o}\'{z}d\'{z}}
\email{andrzej.gozdz@umcs.lublin.pl}
\affiliation{Institute of Physics, Maria Curie-Sk{\l}odowska
University, pl.  Marii Curie-Sk{\l}odowskiej 1, 20-031 Lublin, Poland}

\author{W{\l}odzimierz Piechocki} \email{wlodzimierz.piechocki@ncbj.gov.pl}
\affiliation{Department of Fundamental Research, National Centre for Nuclear
  Research, Ho{\.z}a 69, 00-681 Warszawa, Poland}

\author{Grzegorz Plewa} \email{grzegorz.plewa@ncbj.gov.pl}
\affiliation{Department of Fundamental Research, National Centre for Nuclear
  Research, Ho{\.z}a 69, 00-681 Warszawa, Poland}

\date{\today}

\begin{abstract}
We present the quantum model of the asymptotic dynamics underlying the Belinski-Khalatnikov-Lifshitz (BKL) scenario.
The symmetry of the physical  phase space enables making use of the affine coherent states quantization.
Our results show that quantum dynamics is regular in the sense that during quantum evolution the expectation values
of all considered observables are finite. The classical singularity of the BKL scenario is replaced by the quantum
bounce that presents a unitary evolution of considered gravitational system.  Our results suggest that quantum general
relativity has a good chance to be free from singularities.
\end{abstract}

\pacs{04.60.-m, 04.60.Kz, 0420.Cv}

\maketitle

\tableofcontents

\section{Introduction}

 It is believed that the cosmological and astrophysical singularities predicted by general relativity (GR) can be resolved
at the quantum level.  That has been shown to be the case in the quantization of the simplest singular GR solutions like FRW-type
spacetimes (commonly used in observational cosmology). However, it is an open problem in the general case. Our paper addresses the
issue of possible resolution of a {\it generic} singularity problem  in GR due to quantum effects.

The Belinski, Khalatnikov and Lifshitz (BKL) conjecture is thought to describe a generic solution to the Einstein equations near
spacelike singularity (see, \cite{BKL22,BKL33,Bel} and references therein). Later, it was extended to deal with generic timelike
singularity of general relativity\cite{P1,P2,Parnovsky:2016mdn}. According to the BKL scenario \cite{BKL22,BKL33}, in  the approach
to a space-like singularity neighbouring points decouple and spatial derivatives become negligible in comparison to temporal derivatives.
The conjecture is based on the examination of the dynamics toward the singularity of a Bianchi spacetime, typically Bianchi IX (BIX).
The BKL scenario presents the oscillatory evolution (towards the singularity) entering the phase of chaotic dynamics (see, e.g.,
\cite{Cornish:1996yg,Cornish:1996hx}), followed by approaching the spacelike manifold with diverging curvature and matter field invariants.

The most general scenario is the dynamics  of the {\it non-diagonal} BIX model. However, this dynamics is difficult
to exact treatment. Qualitative analytical considerations \cite{ryan,bel,Jantzen:2001me} and numerical analysis \cite{Nick} strongly suggest
that in the asymptotic regime near the singularity the exact dynamics can be well approximated by much simpler dynamics (presented in the next section).

The BKL scenario based on a {\it diagonal} BIX reduces to the dynamics described in terms of the three directional scale
factors dependent on an evolution parameter (time). This dynamics towards the singularity has the following properties:
(i) is symmetric with respect to the permutation of the scale factors, (ii) the scale factors are oscillatory functions
of time, (iii) the product of the three scale factors is proportional to the volume density decreasing monotonically
to zero, and (iv) the scale factors may intersect each other during the evolution of the system. The diagonal BIX is suitable
to address the vacuum case and the cases with simple matter fields. More general cases,
including perfect fluid with nonzero time dependent velocity, require taking non-diagonal space metric.
However, the general dynamics simplifies near the singularity and can be described by three {\it effective} scale factors,
which include contribution from matter field.
This effective dynamics does not have the properties (i) and (iv) of the diagonal case. More details can be found in
the paper \cite{Czuchry:2014hxa}.

Roughly speaking, the main advantage of the non-diagonal BIX scenario is that it can be used to derive the BKL conjecture
in a much simpler way than when starting from the diagonal case. Namely, considering inhomogeneous perturbations of the
non-diagonal BIX metric is sufficient to derive the BKL conjecture, whereas the diagonal case needs additionally considering
inhomogeneous perturbation of the matter field that would correspond, e.g.,  to time dependent velocity of the perfect fluid
\cite{BielV}.

The present paper concerns the {\it quantum} fate of the asymptotic dynamics of the non-diagonal BIX model.
The quantum dynamics, described by the Schr\"{o}dinger equation, is regular (no divergencies of physical observables)
and the evolution is unitary.  The classical singularity is replaced by quantum bounce due to the continuity of the probability
density.

Our paper is organized as follows: In Sec. \!II we recall  the Hamiltonian formulation of our gravitational system and identify
the topology of physical phase space.
Section III is devoted to the construction of the quantum formalism.  It is based on using the affine coherent states ascribed
to the physical phase space, and the resolution of the unity in the carrier space of the unitary representation of the affine
group.  The quantum dynamics is presented  in Sec. \!IV.  Finding an explicit solution
to the Schr\"{o}dinger equation enables addressing the singularity problem. We conclude in the last section.
Appendix A presents an alternative affine coherent states. The basis of the carrier
space is defined in App. B.

\section{Classical dynamics}

For self-consistency of the present paper, we first recall some results of Ref. \!\cite{Czuchry:2012ad},
followed by the analysis of the topology of the physical phase space.

\subsection{Asymptotic regime of general Bianchi IX dynamics}

The dynamical equations of the general (nondiagonal) Bianchi IX model, in the evolution towards the singularity,
take the following {\it asymptotic} form  \cite{ryan,bel}
\begin{equation}\label{a1}
 \frac{d^2\ln a}{d \tau^2}  = b/a - a^2,~~~\frac{d^2\ln b}{d \tau^2}
= a^2-b/a + c/b,~~~\frac{d^2\ln c}{d \tau^2} =
 a^2 - c/b,
\end{equation}
where $a,b,c$ are functions of an evolution parameter $\tau$, and are interpreted
as the {\it effective} directional scale factors of considered anisotropic universe.  The
solution to \eqref{a1} should satisfy the dynamical constraint
\begin{equation}\label{a2}
\frac{d\ln a}{d \tau}\;\frac{d\ln b}{d \tau}
+ \frac{d\ln a}{d \tau}\;\frac{d\ln c}{d \tau}
+ \frac{d\ln b}{d \tau}\; \frac{d\ln c}{d \tau} - a^2 - b/a - c/b = 0 \, .
\end{equation}
Eqs. \!\eqref{a1} and \eqref{a2} define a coupled highly nonlinear system of ordinary differential equations.

The derivation of this asymptotic dynamics \eqref{a1}--\eqref{a2} from the exact one is based, roughly speaking,  on the assumption
that in the evolution towards the singularity ($\tau \rightarrow 0$) the following conditions are satisfied:
\begin{equation}\label{rat1}
  a \rightarrow 0,~~~b/a  \rightarrow 0,~~~ c/b  \rightarrow 0 \, .
\end{equation}
We recommend Sec. \!6 of Ref.  \cite{bel} for the justification of taking this assumption\footnote{The directional scale factors
$\{a,b,c\}$ considered here and in \cite{ryan}, and the ones considered in \cite{bel} named $\{A,B,C\}$, are connected by the relations:
$a = A^2, b = B^2, c = C^2$.}.  The numerical simulations of the exact dynamics presented in the recent paper \cite{Nick} give
support to the assumption \eqref{rat1} as well.

The dynamics \eqref{a1}--\eqref{a2} defines the asymptotic regime of the BKL scenario, satisfying \eqref{rat1}, that lasts quite
a long time before the system approaches the singularity, marked by the condition
$a\, b\, c \rightarrow 0$ (see \cite{ryan,bel} for more details).

\subsection{Hamiltonian formulation with dynamical constraint}

Roughly speaking,  using the canonical phase space variables $\{q_k,p_l\} = \delta_{kl}$ introduced in \cite{Czuchry:2012ad}:
\begin{equation}\label{hc1}
  q_1 := \ln a,~~~q_2 := \ln b,~~~q_3 := \ln c,~~~\mbox{and}~~~p_1 = \dot{q}_2 + \dot{q}_3,~~~p_2 = \dot{q}_1 + \dot{q}_3,~~~p_3
= \dot{q}_1 + \dot{q}_2 \, ,
\end{equation}
where ``dot'' denotes $d/d\tau$, turns the constraint \eqref{a2} into the Hamiltonian constraint $H_c = 0$ defined by
\begin{equation}\label{hc2}
H_c :=  \frac{1}{2}(p_1 p_2 + p_1 p_3 + p_2 p_3)
- \frac{1}{4} (p_1^2 + p_2^2 + p_3^2) - \exp (2 q_1) - \exp (q_2
-q_1)- \exp (q_3 -q_2) = 0 \, ,
\end{equation}
and the corresponding Hamilton's equations read
\begin{eqnarray}
  \dot{q}_1 &=& \frac{1}{2} (-p_1 + p_2 +p_3), \label{x1}\\
 \dot{q}_2 &=& \frac{1}{2} (p_1 - p_2 + p_3),\label{x2}\\
\dot{q}_3 &=& \frac{1}{2} (p_1 + p_2 - p_3),\label{x3}\\
\dot{p}_1 &=& 2\exp(2 q_1) - \exp (q_2 -q_1), \label{p1}\\
\dot{p}_2 &=& \exp (q_2
-q_1)- \exp (q_3 -q_2), \label{p2}\\
\dot{p}_3 &=& \exp (q_3 -q_2).\label{p3}
\end{eqnarray}
The dynamical systems analysis applied to the system \eqref{hc2}--\eqref{p3} leads to the conclusion
that there exists the set of the nonhyperbolic type of critical points $S_B$, corresponding to this dynamics,
defined by  \cite{Czuchry:2012ad}
\bea
\label{critS}S_B: &=& \{(q_1,q_2,q_3,p_1,p_2,p_3)\in \bar{\dR}^6
~|~ (q_1 \rightarrow
    -\infty,~ q_2-q_1 \rightarrow -\infty,~ q_3-q_2 \rightarrow -\infty)\nonumber
    \\&& \wedge (p_1 = 0 = p_2 = p_3  \label{critS0}\}, \eea
where $\bar{\dR}:= \dR \cup  \{-\infty, +\infty\}$.

Eqs. \!\eqref{critS} and \eqref{rat1} imply that the space of singular points includes the critical points surface $S_B$.
It is so because $(q_1 \rightarrow -\infty,~ q_2-q_1 \rightarrow -\infty,~ q_3-q_2 \rightarrow -\infty)$
implies \eqref{rat1} (which means $a\, b\, c \rightarrow 0$) for any $(p_1, p_2, p_3) \in \bar{\dR}^3$.

\subsection{Hamiltonian formulation devoid of dynamical constraint}

There exists the reduced phase space formalism  corresponding to the dynamics \eqref{hc2}--\eqref{p3} presented in
\cite{Czuchry:2012ad}. The two form $\Omega$ defining the Hamiltonian formulation, devoid of the dynamical constraint
\eqref{hc2}, is given by
\begin{equation}\label{sim1}
\Omega = dq_1 \wedge dp_1 + dq_2 \wedge dp_2 + dt \wedge dH,
\end{equation}
where  $t:= p_3$. The Hamiltonian $H$ is defined to be $H= - q_3 $, where $q_3$ is determined
from the dynamical constraint \eqref{hc2}.
The variables $\{q_1, q_2,p_1, p_2\}$ parameterise the physical phase space, $H= H(t,q_1, q_2,p_1, p_2)$ is the Hamiltonian generating
the dynamics, and $t$ is an evolution parameter  corresponding to the specific  choice of $H$.
The Hamiltonian reads\footnote{In what follows, the choice of $H$ differs from the one presented in \cite{Czuchry:2012ad}
by a factor minus one to fit properly the third term of the r.h.s. of \eqref{sim1}.}
\begin{equation}\label{sim3}
H(t, q_1, q_2, p_1, p_2)
= - q_2 - \ln \left[-e^{2 q_1}-e^{q_2 - q_1}-\frac{1}{4}(p_1^2
+ p_2^2 +t^2) +\frac{1}{2}(p_1 p_2 + p_1 t + p_2 t)\right],
\end{equation}
and Hamilton's equations are
\begin{eqnarray}
\label{ff1}
\frac{d q_1}{d t} &=& \frac{\partial H}{\partial p_1}
=  \frac{  p_1 - p_2 - t }{2 F},\\
\label{ff2}
\frac{d q_2}{d t} &=& \frac{\partial H}{\partial p_2}
=  \frac{- p_1 + p_2 - t }{2 F},\\
\label{ff3}
\frac{d p_1}{d t} &=& - \frac{\partial H}{\partial q_1}
= \frac{-2 e^{2 q_1}+ e^{ q_2 - q_1 }}{F}  ,\\
\label{ff4}
\frac{d p_2}{d t} &=&- \frac{\partial H}{\partial q_2}
= 1 - \frac{e^{q_2 - q_1}}{F},
\end{eqnarray}
where
\begin{equation}\label{eq1}
F(t, q_1,q_2,p_1,p_2):= -e^{2 q_1}-e^{q_2 -q_1}
-\frac{1}{4}(p_1^2 + p_2^2 +t^2)+\frac{1}{2}(p_1 p_2 + p_1 t + p_2 t)> 0 .
\end{equation}

In what follows, we do not use explicitly the relationship between the evolution parameter $\tau$ and $t$, but it does exist. Namely, since
$t = p_3$,   solving the dynamics \eqref{hc2}--\eqref{p3} would give $p_3 = p_3 (\tau)$. Moreover, making use of Eq. \!\eqref{p3} we can get the essential
information on the time variable: (i) $\dot{p}_3 > 0$ so that $t$ is an increasing function of $\tau$, and (ii) integrating \eqref{p3} gives
$p_3 = \int_{\tau_1}^{\tau_2}d\tau\, \exp (q_3 - q_2) > 0$ as the integrand is positive definite.

The reduced system \eqref{ff1}--\eqref{ff4} has been obtained  in the procedure of mapping the
system with the Hamiltonian constraint, defined by \eqref{hc2}--\eqref{p3}, into the Hamiltonian system devoid of the constraint. In the
former, the Hamiltonian $H_c$ is a dynamical constraint, in the latter the Hamiltonian  $H$ in a generator of dynamics without the constraint.
As it is known, this procedure is a sort of  ``one-to-many'' mapping (for more details, see e.g. \cite{Malkiewicz:2017cuw,Malkiewicz:2015fqa}
and references therein). Roughly speaking, it consists in resolving the dynamical constraint with respect to one phase space variable that is
chosen to be a Hamiltonian.
This procedure leads to the choice of an evolution parameter (time) as well so that the  Hamiltonian and time emerge  in a single step.
In general, this is a highly non-unique procedure if there are no hints to this process. Here the choice of the reduction was motivated by the two
circumstances: (i) the resolution of the constraint only with respect to  one  variable $q_3$ is unique, and (ii) the resulting reduced phase space
is isomorphic to the Cartesian product of two affine groups. The latter has unitary irreducible representation enabling the affine coherent states
quantization of the underlying gravitational system (presented in the next section).

\subsection{Numerical simulations of dynamics}

\begin{figure}
\begin{minipage}{1\linewidth}
  \centering
  \subfloat[]{\includegraphics[width=0.50\textwidth]{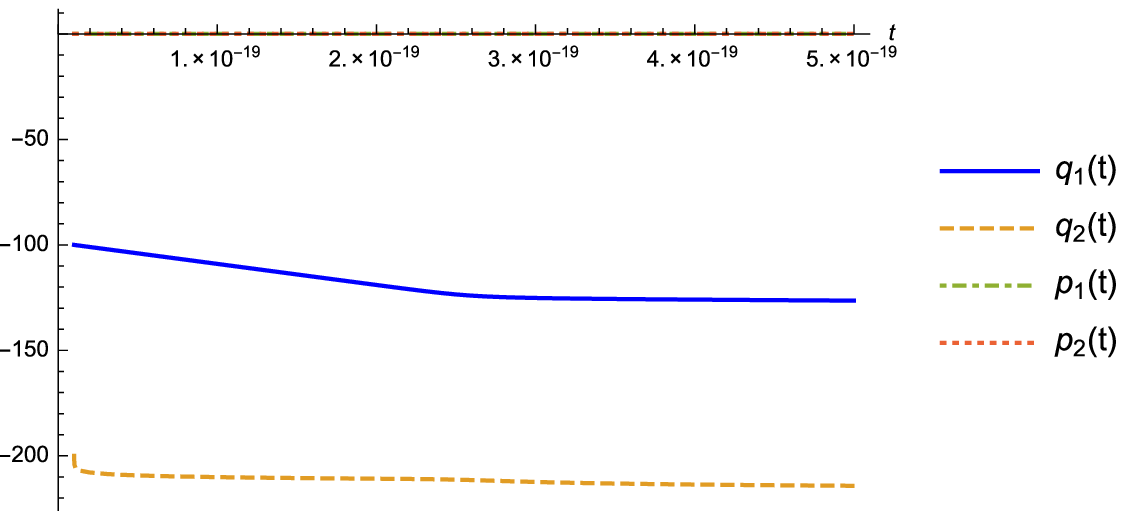}}
  \hfill
  \subfloat[]{\includegraphics[width=0.45\textwidth]{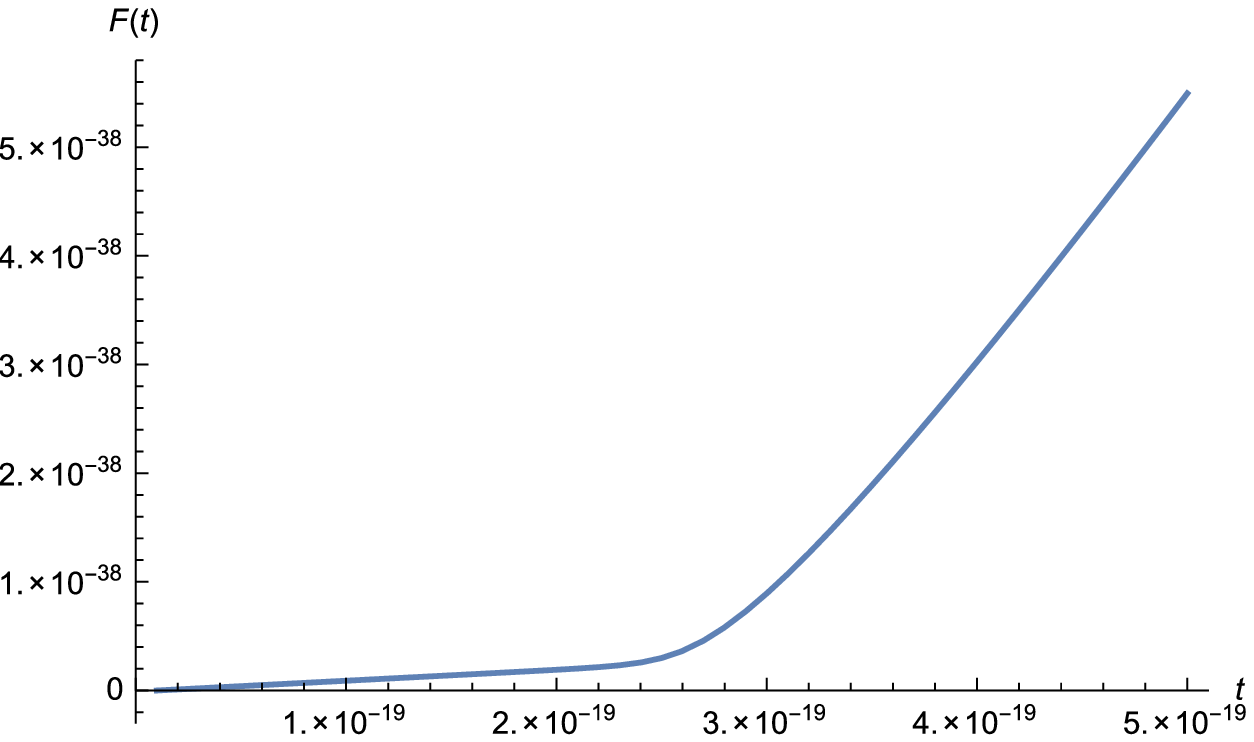}}
\end{minipage}\par\medskip
\begin{minipage}{1\linewidth}
\centering
  \subfloat[]{\includegraphics[width=0.45\textwidth]{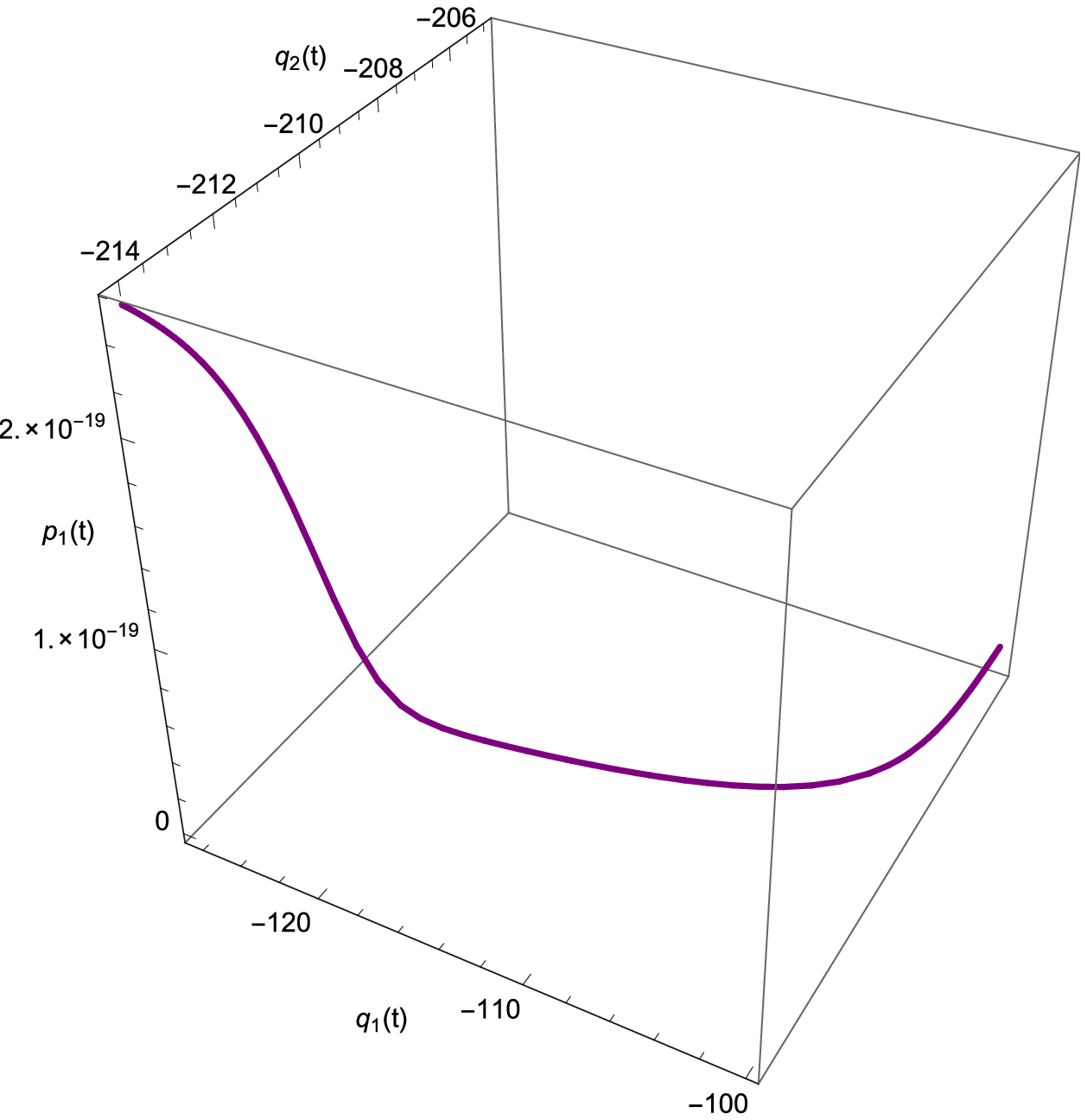}}
  \hfill
  \subfloat[]{\includegraphics[width=0.45\textwidth]{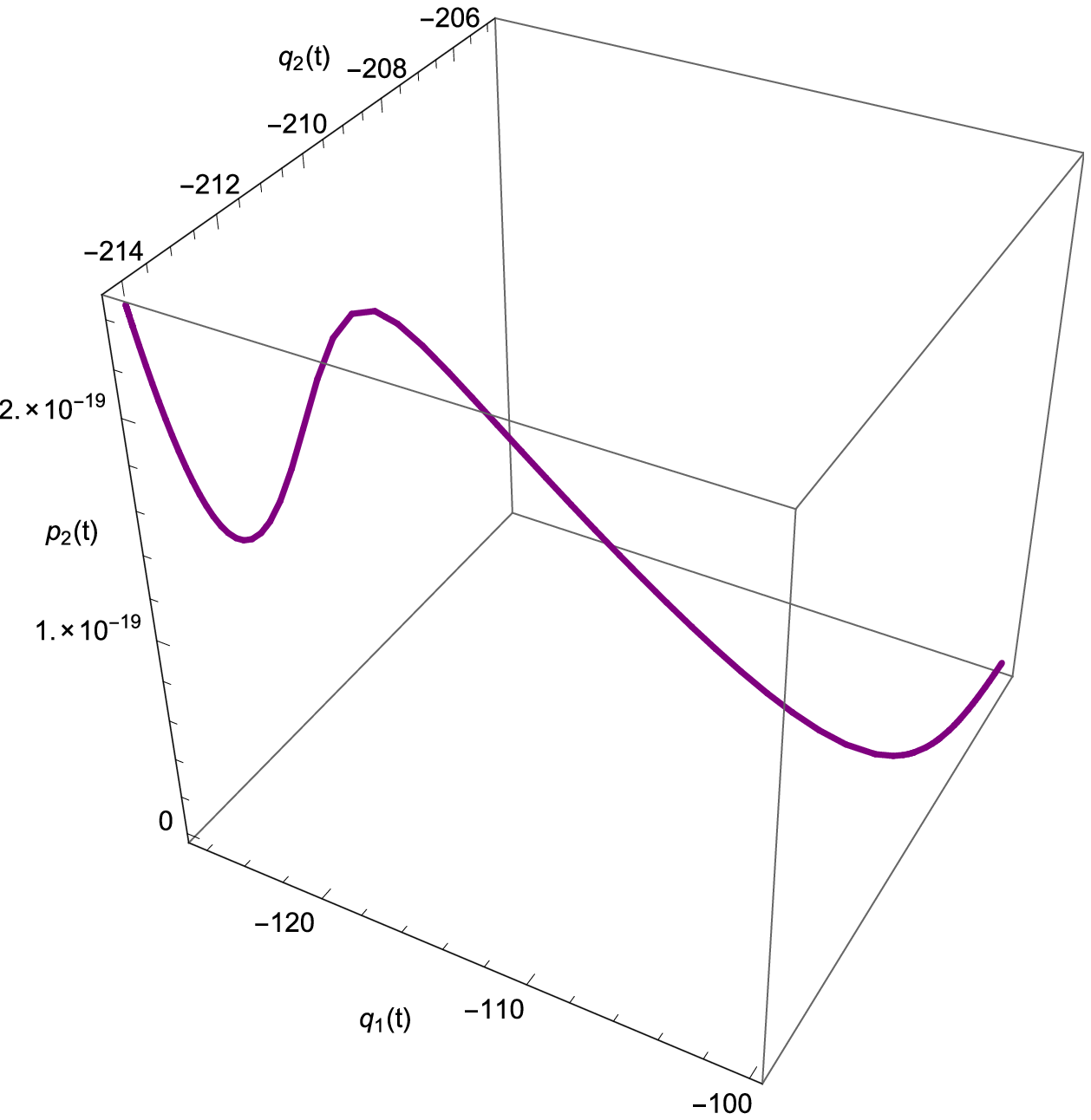}}
\end{minipage}
\caption{The solution to Eqs. \!\eqref{ff1}--\eqref{ff4} with the initial conditions: $\Lambda = 100$, $\delta = t_0 =  10^{-20}$, which corresponds to $q_1 (t_0)=q_2(t_0)-q_1(t_0)=-100$, $\tilde{F}_0 \simeq 3.72 \times 10^{-44}$.}
\label{plot}
\end{figure}

In what follows we present the numerical solutions of Eqs. \!\eqref{ff1}--\eqref{ff4} near the singularity, which corresponds to the case:
$q_1 \rightarrow -\infty$, $q_2-q_1 \rightarrow -\infty$, $F \rightarrow 0^+$. Accordingly, we choose the initial conditions in the form:
\begin{align}
\nonumber
q_1(t_0) = -\Lambda, \quad p_1(t_0) = \frac{1}{2} \left( t_0+\sqrt{t_0^2-2 e^{-\Lambda}(4+\delta)} \right),
\\[1ex]
\label{bound}
 \quad q_2(t_0) = - 2 \Lambda, \quad p_2(t_0) = \frac{1}{2} \left( t_0-\sqrt{t_0^2-2 e^{-\Lambda}(4+\delta)} \right),
\end{align}
where $t_0$ is the initial ``time'', whereas $\Lambda$ and $\delta$ denote two additional parameters. Inserting \eqref{bound} into \eqref{eq1}
one gets
\be
F(t_0, q_1(t_0),q_2(t_0),p_1(t_0),p_2(t_0)) = \left( 1+\frac{\delta}{2} \right) e^{-\Lambda} - e^{-2 \Lambda} =: \tilde{F}_0.
\ee
Thus, taking the limit $\Lambda \rightarrow \infty$, while keeping $\delta$ and $t_0$ to be fixed, one gets
$\tilde{F}_0 \rightarrow 0^+$, $q_1 \rightarrow -\infty$, $q_2-q_1 \rightarrow -\infty$. Therefore,  \eqref{bound} can be regarded as an example
that ensures that we are close to the singularity. In particular, taking $\Lambda \gg 1$ and choosing $\delta>0$ to be small (but not
necessarily $\delta \ll 1$), for a fixed $t_0$, we get the vicinity of the singularity. In the next step we solve  the system
\eqref{ff1}--\eqref{ff4} with the boundary conditions \eqref{bound}. To be more specific, starting with the point in the phase space
defined by \eqref{bound}, we can solve \eqref{ff1}--\eqref{ff4} forward or backward in time. Below, we consider the first possibility
assuming $t \geq t_0$. An example solution is presented in Fig \ref{plot}.
The evolution with $t_0 > t \rightarrow 0$,  becomes quickly meaningless because numerical errors grow
so much that the results cannot be trusted. This is due to rapidly increasing curvature of the
underlying spacetime and increasing nonlinearity effects near the surface of the nonhyperbolic critical
points \eqref{critS}. These lead to faster and faster changes of the phase space variables parameterizing the dynamics,
and finally to breaking off our numerics. We have tested that our choice $t_0 = 10^{-20}$ enables
performing reliable numerical simulations for $t\geq t_0$.

The solution visible in Fig \ref{plot} presents a wiggled curve in the physical phase space. This classical dynamics cannot be
further extended towards the singularity (for $t < t_0$) due to the physical and mathematical reasons (and implied numerical difficulties).

\subsection{Topology of phase space}

Equations
\eqref{ff1}--\eqref{ff4} define a coupled system of nonlinear ordinary
differential equations. The solution defines the  phase space of our
gravitational system.

Applying the simple algebraic identity $(\alpha + \beta + \gamma)^2 = \alpha^2 + \beta^2 + \gamma^2 + 2\alpha\beta +
2\beta\gamma + 2\alpha\gamma$ to Eq. \eqref{eq1} gives:
\begin{align}
F(t, q_{1}, q_{2}, p_{1}, p_{2}) &= -e^{2q_{1}} - e^{q_{2} - q_{1}}
- \frac{1}{4}(p_1 - p_2 + t)^2 + p_1 t , \label{eq2} \\
F(t, q_{1}, q_{2}, p_{1}, p_{2}) &= -e^{2q_{1}} - e^{q_{2} - q_{1}}
- \frac{1}{4}(-p_1 + p_2 + t)^2 + p_2 t \label{eq3}, \\
F(t, q_{1}, q_{2}, p_{1}, p_{2}) &= -e^{2q_{1}} - e^{q_{2} - q_{1}}
- \frac{1}{4}(p_1 + p_2 - t)^2 + p_1 p_2 \label{eq4}.
\end{align}
Combining \eqref{eq1} and \eqref{eq2} we get
\begin{equation}\label{eq5}
p_1 > \frac{1}{t}\left[e^{2q_{1}} + e^{q_{2} - q_{1}}
+ \frac{1}{4}(p_1 - p_2 + t)^2\right] ,
\end{equation}
whereas \eqref{eq1} and \eqref{eq3} give
\begin{equation}\label{eq6}
p_2 > \frac{1}{t}\left[e^{2q_{1}} + e^{q_{2} - q_{1}}
+ \frac{1}{4}(-p_1 + p_2 + t)^2\right] .
\end{equation}
Making use of \eqref{eq1} and \eqref{eq4} leads to
\begin{equation}\label{eq7}
p_1 p_2 > e^{2q_{1}} + e^{q_{2} - q_{1}} +
\frac{1}{4}(p_1 + p_2 - t)^2 .
\end{equation}
It is clear that the signs of both r.h.s. of Eq. \!\eqref{eq5} and
Eq. \!\eqref{eq6} depend only on the sign of $t$. Since $t >0$, we get $p_1 > 0$ and $p_2 > 0$.

To examine the issue of well definiteness of the logarithmic function in \eqref{sim3}, we  rewrite Eq. \!\eqref{sim3} in terms of $(q_k, p_k)$
variables\footnote{We stay with $p_3 = t$ as we wish to discuss the possible sign of the time variable.}
as follows
\begin{equation}\label{sim4}
  q_3 - q_2 = \ln \left[-e^{2 q_1}-e^{q_2 - q_1}-\frac{1}{4}(p_1^2
+ p_2^2 +t^2) +\frac{1}{2}(p_1 p_2 + p_1 t + p_2 t)\right] \, .
\end{equation}
At the critical surface $S_B$, defined by \eqref{critS}, we have $q_3 - q_2 \rightarrow -\infty$
so the r.h.s. of \eqref{sim4} should have this property as well. It means that $F(t,q_1,q_2,p_1,p_2)\rightarrow 0^+$
on approaching $S_B$ so the problem of well definiteness reduces
to solving the equation $t^2 - 2 (p_1 + p_2)t + (p_1 - p_2)^2 < 0$ with respect to the time variable. The solution reads
$(\sqrt{p_1} -  \sqrt{p_2})^2 < t < (\sqrt{p_1} +  \sqrt{p_2})^2 $, which means that as $p_1 \rightarrow 0$ and $p_2 \rightarrow 0$,
we have $t \rightarrow 0^+$.
Therefore, our gravitational system evolves away from the singularity at $t = 0$.

The range of the variables $q_1$ and $q_2$ results from the physical
interpretation ascribed to them \cite{Czuchry:2012ad}. Since $0 <a < +\infty$
and $0 <b < +\infty$, we have $(q_1, q_2)\in \dR^2$.  Thus, the physical phase
space $\Pi$ consists of the two half planes:
\begin{equation}\label{kps2}
\Pi = \Pi_1 \times \Pi_2 :=
\{(q_1, p_1) \in \dR \times \dR_+\} \times
\{(q_2, p_2) \in \dR \times \dR_+\} \; ,
\end{equation}
where $\dR_+ := \{p \in \dR~|~p>0 \}$. Each $\Pi_k~(k=1,2)$ can be identified with the
manifold of the affine group $\Group{Aff}(\dR)$ acting on $\dR$, which is sometimes
denoted as ``$px+q$''
\begin{equation*}
\label{OneDimAffGrp}
x'=(q,p)\cdot x=px+q, \mbox{ where } p>0 \mbox{ and } q \in \dR \, .
\end{equation*}
This opens the possibility for quantization by affine coherent states.

It is important to notice that only the subspace
\begin{equation} \label{AvailablePhaseSpace}
\tilde{\Pi} =\{(q_1,p_1,q_2,p_2)
\colon F(t,q_1,p_1,q_2,p_2)>0 \} \subset \Pi,
\end{equation}
where $\Pi$ defined by \eqref{kps2}, is available to the dynamics. It is due to
the logarithmic function in the expression \eqref{sim3} defining the
Hamiltonian. To make this restriction explicit, we rewrite the Hamiltonian
\eqref{sim3} in the form
\begin{eqnarray} \label{vis1a}
&& H(t, q_1, q_2, p_1, p_2) =
\begin{cases}
- q_2 - \ln F(t, q_1, q_2, p_1, p_2),~~\text{for}~~F(t, q_1, q_2, p_1, p_2) >0 \\
0,~~~\text{for}~~~F(t, q_1, q_2, p_1, p_2) < 0 \,
\end{cases}
\end{eqnarray}
with $\lim_{F\rightarrow 0^-} H = 0~~\text{and}~~\lim_{F\rightarrow 0^+} H = + \infty $.

\section{Quantization}

Suppose we have reduced phase space Hamiltonian formulation of classical
dynamics of a gravitational system. It means dynamical constraints have been
resolved and the Hamiltonian is a generator of the dynamics.  By quantization we
mean (roughly speaking) a mapping of such Hamiltonian formulation into a quantum
system described in terms of quantum observables (including Hamiltonian)
represented by an algebra of  operators acting in a Hilbert space. The
construction of the Hilbert space may make use some mathematical properties of
phase space like, e.g., symplectic structure, geometry or topology.  The quantum
Hamiltonian is used to define the Schr\"{o}dinger equation.  In what
follows we make specific the above procedure by using the affine coherent states
approach.

\subsection{Affine coherent states}

The Hilbert space $\mathcal{H}$ of the entire system consists of the Hilbert
spaces ${\mathcal{H}}_1$ and ${\mathcal{H}}_2$ corresponding to the phase spaces
$\Pi_1$ and $\Pi_2$, respectively. In the sequel the construction of
$\mathcal{H}_1$ is followed by merging of $\mathcal{H}_1$ and
$\mathcal{H}_2$.

As both half-planes $\Pi_1$ and $\Pi_2$ have the same mathematical structure,
the corresponding Hilbert spaces $\mathcal{H}_1$ and $\mathcal{H}_2$ are
identical so we first consider only one of them. In what follows we present the
formalism for $\Pi_1$ and $\mathcal{H}_1$ to be extended later to the entire
system.

\subsubsection{Affine coherent states for half-plane}

The phase space $\Pi_1$ may be identified with the affine group $\Group{G}_1
\equiv \Group{Aff}(\dR)$ by defining the multiplication law as follows
\begin{equation}\label{c1b}
(q^\prime, p^\prime)\cdot (q, p) = (p^\prime q+ q^\prime, p^\prime p ),
\end{equation}
with the unity $(0,1)$ and the inverse
\begin{equation}\label{c2b}
(q^\prime, p^\prime)^{-1} = (-\frac{q^\prime}{p^\prime}, \frac{1}{p^\prime}).
\end{equation}
The affine group has two, nontrivial, inequivalent irreducible unitary
representations \cite{Gel,AK1,AK2}.  Both are realized in the Hilbert space
$\mathcal{H}_1=L^2(\dR_+, d\nu(x))$, where $d\nu(x)=dx/x$ is the invariant
measure\footnote{The general notion of
  invariant measure $dm(x)$ on the set $X$ in respect to the transformation $h:
  X \to X$ can be approximately defined as follows: for every function $f\colon
  X \to \dC$ the integral defined by this measure fulfils the invariance
  condition:
%
\[     \int_X dm(x) f(h(x)) = \int_X dm(x) f(x) \, .\]
%
This property is often written as: $dm(h(x))=dm(x)$.} on the multiplicative group $(\dR_+,\cdot)$.
In what follows we choose one of it defined by the following action:
\begin{equation}\label{im1b}
U(q,p)\psi(x)= e^{i q x} \psi(px)\, ,
\end{equation}
where\footnote{We use Dirac's notation whenever we wish to deal with abstract
  vector, instead of functional representation of the vector.}  $|\psi\rangle
\in L^2(\dR_+, d\nu(x))$. Eq. \!\eqref{im1b} defines the representation as we
have
\begin{equation}\nonumber
U(q',p')[U(q,p,)\psi(x)]=U(q',p')[e^{iqx} \psi(px)]=
 e^{i(p'q+q')x} \psi(p'px) \, ,
\end{equation}
and on the other hand
\begin{equation}\nonumber
[U(q',p')U(q,p,)]\psi(x)= U(p'q+q',p'p) \psi(x) = e^{i(p'q+q')x} \psi(p'px)\, .
\end{equation}
This action is unitary in respect to the scalar product in
$L^2(\dR_+, d\nu(x))$:
\begin{eqnarray} \label{ActionUnitarity}
&& \int_0^\infty d\nu(x) [U(q,p) f_2(x)]^\star [U(q,p) f_1(x)]
= \int_0^\infty d\nu(x) [e^{iqx}f_2(px)]^\star [e^{iqx} f_1(px)] \nonumber \\
&& = \int_0^\infty d\nu(x) f_2(px)^\star f_1(px)=
\int_0^\infty d\nu(x) f_2(x)^\star f_1(x)\, .
\end{eqnarray}
The last equality results from the invariance of the measure
$d\nu(px)=d\nu(x)$.

The affine group is not the unimodular group. The left and right invariant
measures are given by
\begin{equation} \label{LRMeasuresAff}
d\mu_L(q,p)=dq\, \frac{dp}{p^2}
\quad \mbox{ and } \quad
d\mu_R(q,p)=dq\, \frac{dp}{p},
\end{equation}
respectively.

The left and right shifts of any  group $\Group{G}$ are defined
differently by different authors.  Here we adopt the definition from
\cite{Jin-Quan}:
\begin{equation} \label{leftRightShiftDef}
\mathcal{L}^L_h f(g)= f(h^{-1}g)
\quad \mbox{and} \quad
\mathcal{L}^R_h f(g)= f(gh^{-1})
\end{equation}
for a function  $f: \mathrm{G} \to \dC$  and all $g \in \Group{G}$.

For simplicity of notation, let us define integrals over the affine group
$\Group{G}_1 =\Group{Aff}(\dR)$ as:
\begin{equation}\label{HaarIntegrals}
\int_{\Group{G}_1} d\mu_L(q,p)= \frac{1}{2\pi}
\int_{-\infty}^{+\infty} dq \int_{0}^\infty \frac{dp}{p^2}
\quad \mbox{and} \quad
\int_{\Group{G}_1} d\mu_R(q,p)= \frac{1}{2\pi}
\int_{-\infty}^{+\infty} dq \int_{0}^\infty \frac{dp}{p} \, .
\end{equation}
In many formulae it is useful to use shorter notation for points in the phase
space $\xi \equiv (q,p)$ and identify them with elements of the affine group. In
this case the product (\ref{c1b}) is denoted as $\xi' \cdot \xi$.  Depending on
needs we will use both notations.

Fixing the normalized vector $\Ket{\Phi} \in L^2(\dR_+, d\nu(x))$, called the
{\it fiducial} vector, one can define a continuous family of {\it affine}
coherent states $\Ket{q,p} \in L^2(\dR_+, d\nu(x))$ as follows
\begin{equation}\label{im2}
\Ket{q,p} = U(q,p) \Ket{\Phi}.
\end{equation}
As we have two invariant measures, one can define two operators which
potentially can lead to the unity in the space $L^2(\dR_+, d\nu(x))$:
\begin{equation}\label{unityOneTwo}
B_L=\int_{\Group{G}_1} d\mu_L(q,p) \Ket{q,p}\Bra{q,p}
\quad \mbox{and} \quad
B_R=\int_{\Group{G}_1} d\mu_R(q,p) \Ket{q,p}\Bra{q,p} \, .
\end{equation}
Let us check which one is invariant under the action $U(q,p)$ of the affine
group:
\begin{equation} \label{BLInvU}
U(q',p') B_L U(q,p)^\dagger = \int_{\Group{G}_1} d\mu_L(q,p)
\Ket{p'q+q',p'p}\Bra{p'q+q',p'p} \, .
\end{equation}
One needs to replace the variables under integral:
\begin{eqnarray} \label{BLInvU2}
&& \tilde{q}=p'q+q' \quad \mbox{and} \quad \tilde{p}=p'p \\
&& q=\frac{1}{p'}(\tilde{q}-q') \quad \mbox{and} \quad p=\frac{\tilde{p}}{p'}.
\end{eqnarray}
Calculating the Jacobian
$\frac{\partial(q,p)}{\partial(\tilde{q},\tilde{p})}=\frac{1}{(p')^2}$ one gets
\begin{equation} \label{BLInvU3}
d\mu_L(q,p)=\frac{1}{p^2} \frac{1}{(p')^2} d\tilde{q}d\tilde{p}=
\frac{1}{\tilde{p}^2} d\tilde{q}d\tilde{p}= d\mu_L(\tilde{q},\tilde{p})\, .
\end{equation}
The last result proves that
\begin{equation} \label{BLInvU}
U(q',p') B_L U(q,p)^\dagger = \int_{\Group{G}_1}
d\mu_L(\tilde{q},\tilde{p}) \Ket{\tilde{q},\tilde{p}}\Bra{\tilde{q},\tilde{p}}
= B_L \, .
\end{equation}
This also means that $B_R$ is not invariant under the action $U(q,p)$.

The {\it irreducibility} of the representation, used to define the coherent
states \eqref{im2}, enables making use of Schur's lemma \cite{BR}, which leads
to the resolution of the unity in $L^2(\dR_+, d\nu(x))$:
\begin{equation}\label{im4}
\int_{\Group{G}_1}  d\mu_L(q,p) \Ket{q,p}\Bra{q,p} = A_\Phi\;\I \; ,
\end{equation}
where the constant $A_\Phi$ can be determined by using any arbitrary, normalized
vector $\Ket{f} \in L^2(\dR_+, d\nu(x))$:
\begin{equation}\label{im3b}
A_\Phi = \int_{\Group{G}_1}d\mu_L(q,p)\,
\BraKet{f}{q,p}\BraKet{q,p}{f} \, .
\end{equation}
This formula can be calculated by making use of the invariance of the measure:
\begin{eqnarray}\label{im3b2}
&& A_\Phi = \int_{\Group{G}_1}  d\mu_L(q,p) \nonumber \\
&&\times \int_0^\infty d\nu(x')\int_0^\infty d\nu(x)
(f(x')^\star e^{iqx'}\Phi(px'))(e^{-iqx}\Phi(px)^\star f(x)) \nonumber \\
&& = \int_0^\infty \frac{dx'}{x'}\int_0^\infty \frac{dx}{x}
\int_0^\infty \frac{dp}{p^2}
\left[\frac{1}{2\pi}\int_{-\infty}^{+\infty} dq e^{iq(x'-x)}\right]
f(x')^\star f(x) \Phi(px')\Phi(px)^\star \nonumber \\
&& = \int_0^\infty \frac{dx}{x^2} |f(x)|^2
\int_0^\infty \frac{dp}{p^2} |\Phi(px)|^2   \nonumber \\
&& = \left(\int_0^\infty \frac{dx}{x} |f(x)|^2\right)
\left(\int_0^\infty \frac{dp}{p^2} |\Phi(p)|^2 \right)
=\int_0^\infty \frac{dp}{p^2} |\Phi(p)|^2
\end{eqnarray}
because  $\BraKet{f}{f}=1$. Thus, the normalization constant is dependent
on the fiducial vector.
Appendix \ref{alternative} presents alternative affine coherent states.

\subsubsection{Structure of the fiducial vector}

The problem which influences the structure of quantum state space is a
possible degeneration of the space due to specific structure of the fiducial
vector. In the case of quantum states the vectors which differ by a phase factor
represent the same quantum state.
Thus, let us consider the states satisfying the above condition for physically
equivalent state vectors \cite{AP}:
\begin{equation}
 \label{FidVectPhase}
 U(\tilde{q},\tilde{p}) \Phi(x)=
 e^{i\beta(\tilde{q},\tilde{p})} \Phi(x),~~
 \mbox{ where }~~~~
 \beta(\tilde{q},\tilde{p}) \in \dR.
 \end{equation}
 The phase space points $\tilde{\xi}=(\tilde{q},\tilde{p})$ treated as elements
 of the affine group $\Group{Aff}(\dR)$ forms its subgroup $\Group{G}_\Phi$.
 The left-hand side of Eq.~\eqref{FidVectPhase} can be rewritten as:
\begin{equation} \label{FidVectPhase2}
e^{i\tilde{q}x}\Phi(\tilde{p}x)=
e^{i \beta(\tilde{q},\tilde{p})} \Phi(x)\, .
 \end{equation}

 If the generalized stationary group $\Group{G}_\Phi$ of the fiducial vector
 $\Phi$ is a nontrivial group, then the phase space points $(q',p')$ and
 $(\tilde{q},\tilde{p}) \cdot (q',p') = (\tilde{p}q',\tilde{p}p')$ are
 represented by the same state vector $U(q',p') \Phi(x)$, for all
 transformations $(\tilde{q},\tilde{p}) \in \Group{G}_\Phi$. This is due to the
 equality
\begin{equation} \label{FidVectPhas3}
U(q',p') \Phi(x)= U((\tilde{q},\tilde{p}) \cdot (q',p')) \Phi(x).
\end{equation}
In this case, to have a unique relation between phase space and the quantum
states, the phase space has to be restricted to the quotient structure
$\Group{Aff}(\dR)/ \Group{G}_\Phi$. From the physical point of view, in most
cases, this is an undesired property.

How to construct the fiducial vector to have $\Group{G}_\Phi=\{e_G\}$, where
$e_G$ is the unit element in this group?
It is seen that Eq.~\eqref{FidVectPhase2} cannot be fulfilled for $\tilde{q}
\neq 0$, independently of chosen fiducial vector. This suggests that the
generalized stationary group $\Group{G}_\Phi$ is parameterized only by the
momenta $(0,\tilde{p})$, i.e. it has to be a subgroup of the multiplicative
group of positive real numbers, $\Group{G}_\Phi \subseteq (\dR_+,\cdot)$.

On the other hand, Eq.~\eqref{FidVectPhase2} implies that
$|\Phi(\tilde{p} x)|= |\Phi(x)|$ for all $(0,\tilde{p}) \in
\Group{G}_\Phi$. In addition, for the fiducial vectors
$\Phi(x)=|\Phi(x)|e^{i\gamma(x)}$ the phases of these complex functions are
bounded by $0\leq \gamma(x) < 2\pi$. Due to Eq.~\eqref{FidVectPhase2} the phases
$\gamma(x)$ and $\beta(0,\tilde{p})$ have to fulfil the following condition
$\gamma(\tilde{p}x)-\gamma(x)=\beta(0,\tilde{p})$. One of the
solutions to this equation is the logarithmic function $\gamma(x)=  \ln(x)$.

In what follows, to have the unique representation of the phase space as a group
manifold of the affine group, we require the generalized stationary group to be
the group consisted only of the unit element. This can be achieved by the
appropriate choice of the fiducial vector.

The unit operator (\ref{im4}) depends explicitly on the fiducial vector
\begin{equation}
\label{UnitOperPhi}
\I[\Phi]= \frac{1}{A_\Phi}
\int_{\Group{G}_1} d\mu_L(\xi) U(\xi)\Ket{\Phi}\Bra{\Phi}U(\xi)^\dagger
\; ,
\end{equation}
This suggests that the most natural transformation of vectors from the
representation given by the fiducial vector $\Ket{\Phi}$ to the representation
given by another fiducial vector $\Ket{\Phi'}$ can be constructed as the
product of two unit operators $\I[\Phi']\I[\Phi]$.

Let us consider an arbitrary vector $\Ket{\Psi} \in L^2(\dR_+,
d\nu(x))$ and its representation in the space spanned with a help of the
fiducial vector $\Ket{\Phi}$:
\begin{equation}
\label{VectorRepFidVectPhi}
\Ket{\Psi}=\I[\Phi]\Ket{\Psi}= \frac{1}{A_\Phi}
\int_{\Group{G}_1} d\mu_L(\xi)
U(\xi)\Ket{\Phi}\Bra{\Phi}U(\xi)^\dagger\Ket{\Psi}
\end{equation}
The same vector can be represented in terms of another fiducial vector
$\Ket{\Phi'}$:
\begin{equation}
\label{VectorRepFidVectTildePhi}
\Ket{\Psi}=\I{[\Phi'] \Ket{\Psi}}= \frac{1}{A_{\Phi'}}
\int_{\Group{G}_0} d\mu_L(\xi)
U(\xi)\Ket{\Phi'} \Bra{\Phi'} U(\xi)^\dagger\Ket{\Psi}
\end{equation}
However, one can transform the vector (\ref{VectorRepFidVectPhi}) into the
vector (\ref{VectorRepFidVectTildePhi}) using the product of two unit operators:
\begin{eqnarray}
\label{TransRepFVPhiFVTildePhi}
&& \Ket{\Psi}=\I [\Phi']\I[\Phi]\Ket{\Psi}= \frac{1}{A_{\Phi'}}\frac{1}{A_\Phi}
\int_{\Group{G}_1} d\mu_L(q,p)
\int_{\Group{G}_1} d\mu_L(q',p')
U(\xi)\Ket{\Phi'} \nonumber \\
&& \Bra{\Phi'} U(\xi^{-1} \cdot \xi') \Ket{\Phi}
   \Bra{\Phi} U(\xi')^\dagger \Ket{\Psi}
\end{eqnarray}
Thus, the choice of the fiducial vector is formally
irrelevant. However, as we will see later a relation between the classical model
and its quantum realization depends on this choice. The sets of affine coherent
states generated from different fiducial vectors may be not unitarily
equivalent, but lead in each case to acceptable affine representations of the
Hilbert space \cite{JRK}.

\subsubsection{Phase space and quantum state spaces}

The quantization procedure requires understanding the relations among the
classical phase space and quantum states space. We have three spaces to be
considered:
\begin{itemize}
\item The phase space $\Pi$, which consists of two half-planes $\Pi_1$ and
  $\Pi_2$ defined by \eqref{kps2}. It is the background for the classical
  dynamics\footnote{For simplicity we consider here only one half-plane, but the
    results can be easily extended to $\Pi$.}.
\item The carrier spaces $ \mathcal{H}_1 :=L^2(\dR_+,
  d\nu(x))$ of the unitary representation $U(q,p)$, with the scalar product
  defined as
\begin{equation}
\label{ScalarProdKX}
\BraKet{\psi_2}{\psi_1}
= \int_0^\infty \frac{dx}{x}\, \psi_2^\star (x) \psi_1(x)\, .
\end{equation}
\item The  space of square integrable functions on the
  affine group $\mathcal{K}_G = L^2(\Group{Aff}(\dR), d\mu_L(q,p))$.
  The scalar product is defined as follows
\begin{equation}
\label{ScalarProdKG}
\BraKet{\psi_{G2}}{\psi_{G1}}_{G}
= \frac{1}{A_\phi}\int_{\Group{Aff}} d\mu_L(q,p)
\psi_{G2}^\star(q,p) \psi_{G1}(q,p)\, ,
\end{equation}
where $\psi_G(q,p):=\BraKet{q,p}{\psi}=\Bra{\Phi}U(q,p)^\dagger\Ket{\psi}$ with
$\Ket{\psi} \in \mathcal{H}_1$. The Hilbert space $\mathcal{K}_G $ is defined to
be the completion in the norm induced by \eqref{ScalarProdKG} of the span of the
$\psi_G$ functions.
\end{itemize}
We show below that the spaces $\mathcal{H}_1$  and $\mathcal{K}_G$
are unitary isomorphic.  First, one needs to check that the functions $ \psi_G
\in \mathcal{K}_G$ are square integrable function belonging to
$L^2(\Group{Aff}(\dR), d\mu_L(q,p))$. Using the decomposition of unity we get
\begin{equation}
\label{L2GSquareIntegFun}
\frac{1}{A_\Phi} \int_{\Group{G}_1} d\mu_L(q,p)|\BraKet{q,p}{\psi}|^2
< \BraKet{\psi}{\psi}_{\mathcal{H}_1} < \infty.
\end{equation}
The definition of the space $\mathcal{K}_G$ shows that for every
$\psi_1,\psi_2 \in \mathcal{H}_l$ we have the corresponding functions
$\psi_{G1},\psi_{G2}$ for which the scalar products are equal (unitarity of the
transformation between both spaces): $\BraKet{\psi_2}{\psi_1}_{\mathcal{H}_1}=
\BraKet{\psi_{G1}}{\psi_{G2}}_{\mathcal{K}_G}$.

Let us now denote by $\Ket{e_n}$ the orthonormal basis in $\mathcal{H}_1$ (see
App. \!\ref{basis}). The corresponding functions $e_{Gn}(q,p)
=\BraKet{q,p}{e_n}$ furnish the orthonormal set:
\begin{eqnarray}
\label{L2GBasis}
&& \BraKet{e_{Gn}}{e_{Gm}}_{\mathcal{K}_G}
=\frac{1}{A_\Phi} \int_{\Group{G}_1} d\mu_L(q,p)
e_{Gn}^\star (q,p)  e_{Gm}(q,p) \nonumber \\
&& = \frac{1}{A_\Phi} \int_{\Group{G}_1} d\mu_L(q,p)
\BraKet{e_n}{q,p}  \BraKet{q,p}{e_m}
= \BraKet{e_n}{e_m}=\delta_{nm} \, .
\end{eqnarray}
It is obvious that the vectors $\Ket{e_{Gn}}$ define the orthonormal basis in
the space $\mathcal{K}_G$. For every vector $\Ket{\psi} \in \mathcal{H}_1$
\begin{equation}
\label{BasisKX}
\Ket{\psi}= \sum_n \BraKet{e_n}{\psi} \Ket{e_n}.
\end{equation}
Closing both sides of the above equation with $\Bra{q,p}$ gives the unique
decomposition of the vector $\psi_G(q,p) \equiv \BraKet{q,p}{\psi} \in
\mathcal{K}_G $ in the basis $\Ket{e_{Gn}}_G$:
\begin{equation}
\label{BasisKX2}
\psi_G(q,p) \equiv \BraKet{q,p}{\psi}
= \sum_n \BraKet{e_n}{\psi} \BraKet{q,p}{e_n}
= \sum_n \BraKet{e_n}{\psi} \Ket{e_{Gn}}_G.
\end{equation}
Note that the vector $\Ket{\psi} \in \mathcal{H}_1$ and the vector $\Ket{\psi_G}
\in \mathcal{K}_G$ have the same expansion coefficients in the corresponding
bases. This define the unitary isomorphism between both spaces. It means that we
can work either with the quantum state space represented by the space
$\mathcal{H}_1$ or $\mathcal{K}_G$.

\subsubsection{Affine coherent states for the entire system}

The phase space $\Pi$ of our classical system has the structure of the Cartesian
product of the two  phase spaces: $\Pi=\Pi_1 \times
\Pi_2$.  The partial phase spaces $\Pi_l$, where $l=1,2$, are identified with
the corresponding affine groups which we denote by
$\Group{G}_l=\Group{Aff}_l(\dR)$. The simple product of both affine groups
$\Group{G}_{\Pi}= (\Group{G}_1=\Group{Aff}_1(\dR)) \times
(\Group{G}_2=\Group{Aff}_2(\dR)) $ can be identified with the whole phase space
$\Pi$:
\begin{equation}
\label{AffActPi}
(\xi_1,\xi_2) \to \Ket{\xi_1,\xi_2 } = U(\xi_1,\xi_2)\Ket{\Phi}
:= U_1(\xi_1) \otimes U_2(\xi_2) \Ket{\Phi},
\end{equation}
where $\xi_l=(q_l,p_l)$, $l=1,2,$ the fiducial vector $\Ket{\Phi}$ belongs to
the simple product of two Hilbert spaces $(\mathcal{H}_1=L^2(\dR_+,d\nu(x_1)))
\times (\mathcal{H}_2=L^2(\dR_+,d\nu(x_2))) = L^2(\dR_+ \times
\dR_+,d\nu(x_1,x_2))$, and where the measure
$d\nu(x_1,x_2)=d\nu(x_1)d\nu(x_2)$. The scalar product in $\mathcal{H}=L^2(\dR_+
\times \dR_+,d\nu(x_1,x_2))$ reads
\begin{equation}
\label{ScalarProdL2Whole}
\BraKet{\psi_2}{\psi_1}= \int_0^\infty d\nu(x_1) \int_0^\infty d\nu(x_2)
\psi_1(x_1,x_2)^\star \psi_2(x_1,x_2) \, .
\end{equation}
The fiducial vector $\Phi(x_1,x_2)$ is constructed as a product of two fiducial
vectors $\Phi(x_1,x_2)= \Phi_1(x_1)\Phi_2(x_2)$ generating the appropriate
quantum partners for the phase spaces $\Pi_1$ and $\Pi_2$. The fiducial vector
of this type does not add any correlations between both partial phase
spaces.  A nonseparable form of $\Phi(x_1,x_2)$ might lead to reducible
representation of $G_\Pi$ in which case Schur's lemma could not be applied to
get the resolution of unity in $\mathcal{H}$.

Let us denote by $\UnitOp_{12}$ the linear extension of the tensor
product $\UnitOp_{1} \otimes \UnitOp_{2}$, where the unit operators in
$\mathcal{H}_k$ are expressed in terms of the appropriate coherent states
\begin{equation}
\label{UnitOperSpacel}
\UnitOp_{k}= \frac{1}{A_{\Phi_k}}
\int_{\Group{G}_k}  d\mu_L(\xi_k)
U_k(\xi_k)\Ket{\Phi_k}\Bra{\Phi_k}U_k(\xi_k)^\dagger,~~~~k = 1,2.
\end{equation}
Let us consider the orthonormal basis $\{e^{(1)}_n(x_1) \otimes
e^{(2)}_n(x_2)\}$ in the Hilbert space $\mathcal{H}$ and an arbitrary vector
$\Psi(x_1,x_2)= \sum_{nm} a_{nm} e^{(1)}_n(x_1) \otimes e^{(2)}_m(x_2)$
belonging to this space (where the basis $e_n(x)$ is defined in
App. \ref{basis}).  Acting on this vector with the operator $\UnitOp_{12}$ one
gets:
\begin{eqnarray}
\label{ActUnitOp12Vect}
\UnitOp_{12} \Psi(x_1,x_2)= \sum_{nm} a_{nm} (\UnitOp_{1}e^{(1)}_n(x_1)) \otimes
(\UnitOp_{2} e^{(2)}_m(x_2))= \Psi(x_1,x_2).
\end{eqnarray}
The operator $\UnitOp_{12}$ is identical with the unit operator $\UnitOp$ on the
space $\mathcal{H}$.

The explicit form of the action of the group
$\Group{G}_{\Pi}$ on the vector $\Psi(x_1,x_2)$ reads:
\begin{eqnarray}
\label{ActGPiH}
&& U(q_1,p_1,q_2,p_2)\Psi(x_1,x_2)=
\sum_{nm} a_{nm} \{U_1(q_1,p_1)e^{(1)}_n(x_1)\}
\otimes \{U(q_2,p_2) e^{(2)}_m(x_2)\} \nonumber \\
&& =\sum_{nm} a_{nm} \{e^{iqx_1} e^{(1)}_n(p_1 x_1)\}
\otimes \{e^{iqx_2} e^{(2)}_m(p_2x_2)\}
=  e^{iq_1x_1} e^{iq_2x_2} \sum_{nm} a_{nm} e^{(1)}_n(p_1 x_1)
\otimes e^{(2)}_m(p_2x_2) \nonumber \\
&& = e^{iq_1x_1} e^{iq_2x_2} \Psi(p_1 x_1,p_2 x_2)\, .
\end{eqnarray}

\subsection{Quantum observables}

Making use of the resolution of the identity \eqref{im4}, we define the
quantization of a classical observable $f$ on a half-plane as follows \cite{Ber}
\begin{equation}\label{im8}
  \mathcal{F} \ni f \longrightarrow  \hat{f} :=
\frac{1}{A_\Phi}\int_{\Group{G}_1} d\mu_L(q,p)
\Ket{q,p} f(q,p) \Bra{q,p}  \in \mathcal{A} \, ,
\end{equation}
where $\mathcal{F}$ is a vector space of real continuous functions on a phase
space, and $\mathcal{A}$ is a vector space of operators (quantum observables)
acting in the Hilbert space $\mathcal{H}_1=L^2(\dR_+, d\nu(x))$. It is clear
that \eqref{im8} defines a linear mapping and the observable $\hat{f}$ is a {\it
  symmetric} (Hermitian) operator. Let us evaluate the norm of the operator
$\hat{f}$:
\begin{equation}
\label{OperNormQuantOper}
\Vert\hat{f}\Vert \leq
\frac{1}{A_\Phi}\int_{\Group{G}_1} d\mu_L (q,p)
|f(q,p)|  \Vert \Ket{q,p}\Bra{q,p} \Vert
\leq \frac{1}{A_\Phi}\int_{\Group{G}_1} d\mu_L(q,p) |f(q,p)| \, .
\end{equation}
This implies that, if the classical function $f$ belongs to the space of
integrable functions $L^1(\Group{Aff(\dR)}, d\mu_L(q,p))$, the operator
$\hat{f}$ is bounded so it is a {\it self-adjoint} operator.  Otherwise, it is
defined on a dense subspace of $L^2(\dR_+, d\nu(x))$ and its possible
self-adjointness becomes an open problem as symmetricity does not assure
self-adjointness so that further examination is required \cite{Reed}.  The
quantization \eqref{im8} can be applied to any type of observables including
non-polynomial ones, which is of primary importance for us due to the functional
form of the Hamiltonian \eqref{sim3}.

It is important to indicate that in the case the classical observable $f$ is defined
only on a subspace of the full phase space, the corresponding quantum operator $\hat{f}$
obtained via \eqref{im8} acts in the entire Hilbert space corresponding to the full
phase space. This is the peculiarity of the coherent states quantization \cite{JRK5}.
Roughly speaking, it results from the non-zero overlap, $\langle q,p|q^\prime,p^\prime\rangle \neq 0$,
of any two coherent states of the entire Hilbert space.
In particular, the classical Hamiltonian \eqref{vis1a} is defined on the subspace $\tilde{\Pi}$ of $\Pi$.
Hovever, the corresponding quantum Hamiltonian \eqref{Qd1} acts in the Hilbert space
$L^2(\dR_+ \times \dR_+,d\nu(x_1,x_2))$ corresponding to $\Pi$.

It is not difficult to show that the mapping \eqref{im8} is
covariant in the sense that one has
\begin{equation}\label{cov}
  U(\xi_0) \hat{f} U^\dag (\xi_0) =
\frac{1}{A_\Phi}\int_{\Group{G}_1} d\mu_L(\xi)
\Ket{\xi} f(\xi_0^{-1}\cdot \xi) \Bra{\xi}
=  \widehat{\mathcal{L}^L_{\xi_0}f} \, ,
\end{equation}
where $\mathcal{L}^L_{\xi_0} f(\xi) = f(\xi_0^{-1}\cdot \xi)$ is the left shift
operation (\ref{leftRightShiftDef}) and $\xi_0^{-1}\cdot \xi=
(q_0,p_0)^{-1}\cdot (q,p) = (\frac{q-q_0}{p_0},\frac{p}{p_0})$.

The mapping \eqref{im8} extended to the Hilbert space $\mathcal{H}=L^2(\dR_+
\times \dR_+,d\nu(x_1,x_2))$ of the entire system and applied to an observable
$\hat{f}$ reads
\begin{equation}\label{QuantObservPi}
 \hat{f}(t)= \frac{1}{A_{\Phi_1} A_{\Phi_2}}\int_{G_\Pi}  d\mu_L(\xi_1,\xi_2)
  \Ket{\xi_1,\xi_2}f(\xi_1,\xi_2)\Bra{\xi_1,\xi_2}\, ,
\end{equation}
where $ d\mu_L(\xi_1,\xi_2) := d\mu_L(q_1,p_1)d\mu_L(q_2,p_2)$.

\section{Quantum dynamics}

The mapping \eqref{QuantObservPi} applied to the classical Hamiltonian
\eqref{sim3} reads
\begin{equation}\label{Qd1}
 \hat{H}(t)= \frac{1}{A_{\Phi_1} A_{\Phi_2}}\int_{G_\Pi}  d\mu_L(\xi_1,\xi_2)
  \Ket{\xi_1,\xi_2}H(t, \xi_1,\xi_2)\Bra{\xi_1,\xi_2}\, ,
\end{equation}
where $t$ is an evolution parameter of the classical level.

The quantum evolution of our gravitational system is defined by the
Schr\"{o}dinger equation:
\begin{equation}\label{Qd2}
  i  \frac{\partial}{\partial s}|\Psi (s) \rangle
= \hat{H}(t) |\Psi (s) \rangle \; ,
\end{equation}
where $|\Psi \rangle \in \mathcal{H}$, and where $s$ is an evolution
parameter at the quantum level.

In general, the parameters $t$ and $s$ are different. To get the consistency
between the classical and quantum levels we postulate that $t = s$, which
defines the {\it time} variable at both levels. It is worth to mention that so
defined time changes monotonically due to the special choice of the parameter
$t$ at the classical level (see the paragraph below \eqref{eq1}). This way we
support the interpretation that Hamiltonian is the generator of classical and
corresponding quantum dynamics.

Near the gravitational singularity, the terms $\exp(2q_1)$ and $\exp(q_2 - q_1)$
in the function $F$ can be neglected, see Eqs. \!\eqref{eq1} and \eqref{critS},
so that we have
\begin{equation}\label{ns1}
  F(t, q_1,q_2,p_1,p_2) \longrightarrow F_0 (t, p_1,p_2) := p_1 p_2-\frac{1}{4} (t-p_1-p_2)^2 \, .
\end{equation}
This form of $F$ leads to the simplified  form of the Hamiltonian \eqref{vis1a} which now reads
\begin{eqnarray} \label{ns2a}
&& H_0 (t,q_2,p_1, p_2) :=
\begin{cases}
- q_2 - \ln F_0(t,p_1, p_2),~~~
\text{for}~~~F_0(t,p_1, p_2) >0 \\
0,~~~\text{for}~~F_0(t,p_1, p_2) < 0
\end{cases}
\end{eqnarray}
with $\lim_{F_0\rightarrow 0^-} H_0 = 0~~\text{and}~~\lim_{F_0\rightarrow 0^+} H_0 = + \infty $.
In fact, the condition
\begin{equation}\label{regF3a}
  F_0 (t, p_1,p_2) > 0
\end{equation}
defines the available part of the physical phase space $\Pi$ for the classical
dynamics, defined by \eqref{kps2}, which corresponds to the approximation
\eqref{ns1}.  Eqs. \!\eqref{ns1}--\eqref{ns2a} define the approximation to our
original Hamiltonian system, defined by Eqs. \!\eqref{sim3}--\eqref{eq1}, to
describe the dynamics in the close vicinity of the singularity.

After long, otherwise straightforward, calculations we get the Schr\"{o}dinger equation \eqref{Qd2} in the form
\be \label{ns3}
 i \frac{\partial}{\partial t} \Psi(t,x_1,x_2) = \hat{H_0}\; \Psi(t,x_1,x_2),
\ee
where
\begin{equation}\label{ns4}
  \hat{H_0}:= i \frac{\partial}{\partial x_2} - i \frac{B}{x_2} -  K(t, x_1, x_2) \, ,
\end{equation}
and where $\Psi(t,x_1,x_2):= \langle x_1, x_2 | \Psi (t) \rangle$. The functions $B$ and $K$ are defined to be
\begin{equation}\label{ns5}
  B := A_{\Phi_2} \int_0^\infty \frac{d p}{p}\, \frac{\partial{\Phi_2 (p)}}{\partial p}\Phi_2^\ast (p)\, ,
\end{equation}
and
\be \label{ns6}
 K(t, x_1, x_2) := \frac{1}{A_{\Phi_1} A_{\Phi_2}}\; \int_0^\infty \frac{d p_1}{p_1^2} \int_0^\infty \frac{d p_2}{p_2^2}
\widetilde{\ln}\big(F_0(t, \frac{p_1}{x_1},\frac{p_2}{x_2})\big) |\Phi_1(p_1 )|^2 |\Phi_2(p_2 )|^2 \, ,
\ee
where
\begin{eqnarray} \label{tildeln}
&& \widetilde{\ln}\big(F_0(t, \frac{p_1}{x_1},\frac{p_2}{x_2})\big) :=
\begin{cases}
\ln\big(F_0(t, \frac{p_1}{x_1},\frac{p_2}{x_2})\big),~~~
\text{for}~~~ F_0(t, \frac{p_1}{x_1},\frac{p_2}{x_2})>0 \\
0,~~~\text{for}~~~F_0(t, \frac{p_1}{x_1},\frac{p_2}{x_2}) < 0
\end{cases}
\end{eqnarray}
with $\lim_{F_0\rightarrow 0^-} \widetilde{\ln}F_0 = 0~~\text{and}~~\lim_{F_0\rightarrow 0^+} \widetilde{\ln}F_0 = - \infty$.

Thus,  $B$ and  $K$ become known after the specification of the fiducial vectors $|\Phi_1 \rangle$ and $|\Phi_2 \rangle$.
It results from the definition of $B$ that one has $B^\ast = 1 - B$. The requirement of $\hat{H_0}$ being Hermitian, leads to the result that
$\Phi_2 (x) \in \dR$ and   $B = 1/2$. However, these results can be obtained in the case the following conditions are satisfied:
\begin{equation}\label{ns7}
\Phi_2 (x) =: x \tilde{\Phi}(x),~~~ \lim_{x \rightarrow 0^+}\tilde{\Phi}(x) = 0,~~~\lim_{x \rightarrow +\infty}\tilde{\Phi}(x) = 0\, ,
\end{equation}
and
\begin{equation}\label{ns8}
\Psi (t, x_1,x_2) =: \sqrt{x_2} \,\tilde{\Psi}(t, x_1,x_2),~~~\lim_{x_2 \rightarrow 0^+}\tilde{\Psi}(t, x_1,x_2) = 0,
~~~\lim_{x_2 \rightarrow +\infty}\tilde{\Psi}(t, x_1,x_2) = 0 \, .
\end{equation}
Due to the above, the equation \eqref{ns3} reduces to the equation
\begin{equation}\label{ns9}
i \frac{\partial}{\partial t} \Psi(t,x_1,x_2) =
 \left( i \frac{\partial }{\partial x_2}  -\frac{i}{2x_2} - K(t,x_1,x_2)
 \right) \Psi(t,x_1,x_2)\, .
\end{equation}

The general solution to Eq. \!\eqref{ns9} is found to be
\begin{equation}\label{ns10}
\Psi(t,x_1,x_2)=\eta(x_1,x_2+t-t_0)\, \sqrt{\frac{x_2}{x_2+t-t_0}}
\, \exp\left(i \int_{t_0}^t K(t',x_1,x_2+t-t')\,dt' \right) \, ,
\end{equation}
where $ t \ge t_0$, and where  $\eta (x_1, x_2):= \Psi (t_0, x_1,x_2)$ is the initial state.

In what follows, we extend the range of the  time variable to include $t_0 = 0$  (we
quantize the  sector $t>0$ of classical dynamics).

To get insight into the meaning of the solution \eqref{ns10}, let us consider
the inner product
\begin{equation} \label{nor1}
\BraKet{\Psi(t)}{\Psi(t)} = \int_{\dR_+^2} d\nu(x_1,x_2)
\frac{x_2 |\eta(x_1,x_2+t)|^2}{x_2+t} =
 \int_0^\infty \frac{dx_1}{x_1}  \int_t^\infty d{x_2}
\frac{ |\eta(x_1,x_2)|^2}{x_2} \,.
\end{equation}
Thus, the norm of the solution decreases in time. To get an unitary evolution, we choose
the initial state in the form
\begin{equation}\label{nor2}
 \eta(x_1,x_2) = 0~~~~\mbox{for}~~~~x_2 < t_H \, ,
\end{equation}
where $t_H > 0$ is a parameter of our model. This condition is consistent with \eqref{ns8} and for $t < t_H$ gives
\begin{equation} \label{nor3}
\BraKet{\Psi(t)}{\Psi(t)} = \int_0^\infty \frac{dx_1}{x_1} \big( \int_{t}^{t_H} + \int_{t_H}^\infty\big) d{x_2}
\frac{ |\eta(x_1,x_2)|^2}{x_2} =
\int_0^\infty \frac{dx_1}{x_1}  \int_{t_H}^\infty d{x_2}
\frac{ |\eta(x_1,x_2)|^2}{x_2} \,,
\end{equation}
which is time independent so that the quantum evolution is unitary.

The condition \eqref{nor2}, which restricts $x_2 \sim 1/p_2$ , see
  \eqref{pw1} and the text below it, implies that the probability of finding the system in the region
  with $x_2 < t_H$ vanishes.

\subsection{Elementary quantum observables}

The affine coherent states quantization procedure introduces the carrier space
$L^2(\dR_+^2, d\nu(x_1,x_2))$. The elementary quantum observables, $\hat{q}_k$
and $\hat{p}_k$, can be expressed in terms of $x_k \in \dR_+~$ ($k=1,2$).
Namely, it is not difficult to find that
\begin{eqnarray}\label{pw1}
 && \hat{p_k} \,\psi (t, x_1,x_2) = \Big(\frac{1}{ A_{\Phi_k}}\int_0^\infty
 \frac{dp_k}{p_k^2}\, p_k \,|\Phi(p_k)|^2\Big)\,\frac{1}{x_k}\, \psi (t,
 x_1,x_2) \nonumber \\
&& = \frac{1}{ A_{\Phi_k}} \frac{1}{x_k}\, \psi (t,
 x_1,x_2),~~~~~k=1, 2 \, ,
\end{eqnarray}
where the fiducial vector $\Phi(p_k)$ is normalized to unity.  The
  multiplication operators representing the generalized momenta \eqref{hc1}
  determine the physical meaning of the variables $x_k$ used in the description of
  quantum states, as proportional to the inverse of the generalized momenta.
Similarly,
\begin{equation}\label{qw1}
  \hat{q_k} \,\psi (t, x_1,x_2) = (-i \frac{\partial}{\partial x_k} + \frac{i}{2x_k})\,
  \psi (t, x_1,x_2),~~~~~k=1, 2 \, ,
\end{equation}
where $\psi \in L^2(\dR_+^2, d\nu(x_1,x_2))$. One can show that both, $\hat{q_k}$ and $\hat{p_k}$, are symmetric
operators on the subspace of this Hilbert space, which consists of the functions satisfying
\begin{equation}\label{symqp}
\lim_{x_1 \to 0^+} \frac{1}{\sqrt{x_1}} \psi(x_1,x_2) = 0 = \lim_{x_2 \to 0^+} \frac{1}{\sqrt{x_2}} \psi(x_1,x_2) \, .
\end{equation}

\subsection{Singularity of dynamics}

According to Sec. \!II, the singularity of the classical dynamics is defined by the conditions:
\begin{equation}\label{qw8}
  q_1 \rightarrow - \infty,~~~q_2 - q_1 \rightarrow -\infty,~~~F_0 \rightarrow 0^+~~~\mbox{as}~~~t \rightarrow 0^+.
\end{equation}

It means that the singularity may only occur at $t = 0$, and one cannot see the reason for the classical dynamics of not
being regular for $t>0$.

If for the $\Psi$ satisfying the Schr\"{o}dinger equation \eqref{ns9} we get
\begin{equation}\label{qw9}
   \lim_{t \rightarrow 0^+}\langle\Psi (t)| \hat{q}_1 |\Psi (t)\rangle = - \infty,~~~
    \lim_{t \rightarrow 0^+}\langle\Psi(t)|\hat{q}_2 - \hat{q}_1|\Psi (t)\rangle = -\infty \, ,
\end{equation}
and in addition
\begin{equation}\label{qw99}
 \lim_{t \rightarrow 0^+}\langle\Psi (t)| \,\hat{F}_0^{(\theta)} |\Psi (t)\rangle  = 0 \, ,
\end{equation}
our quantization fails in resolving the singularity problem of the classical dynamics\footnote{The precise meaning of Eq. \!\eqref{qw99}
will become clear in the next subsection.}.

The operator $\hat{F}^{(\theta)}_0$, which occurs in  \eqref{qw99},  is of basic importance and is found to be
\begin{equation}\label{regF1}
\hat{F}^{(\theta)}_0 (t)\,\Psi (t, x_1,x_2) = \breve{F}^{(\theta)}_0 (t,x_1,x_2)\,\Psi (t, x_1,x_2) \, ,
\end{equation}
where
\begin{equation}\label{regF2}
\breve{F}^{(\theta)}_0 (t,x_1,x_2) :=    \frac{1}{A_{\Phi_1} A_{\Phi_2}}\; \int_0^\infty \frac{d p_1}{p_1^2} \int_0^\infty \frac{d p_2}{p_2^2}
F_0^{(\theta)}(t, \frac{p_1}{x_1},\frac{p_2}{x_2}) |\Phi_1(p_1 )|^2 |\Phi_2(p_2 )|^2 \, ,
\end{equation}
with
\begin{equation}\label{thetka}
  F_0^{(\theta)}(t, \frac{p_1}{x_1},\frac{p_2}{x_2}) := \theta \big( F_0(t, \frac{p_1}{x_1},\frac{p_2}{x_2})\big)\;F_0(t, \frac{p_1}{x_1},\frac{p_2}{x_2})\, ,
\end{equation}
where $\theta$ is the Heaviside step function defined  as follows: $\theta (x) = 1$ for $x > 0$ and $\theta (x) = 0$ for $x \leq 0$.
Therefore, $\hat{F}^{(\theta)}_0$ is a multiplication operator.

\subsection{Resolution of the singularity problem}

In what follows, we first define an example of a regular state at $t = t_s >0$.
Next, we make  generalization. Afterwards, we map the general regular state to the  initial state at $t = 0 $, by inverting the general form
of the solution defined by Eqs. \!\eqref{ns10} (with $t_0 = 0$) and \eqref{nor2}. Finally, we argue that the initial state is regular at $t = 0$ due
to the unitarity of the quantum evolution. This way we get the resolution of the initial singularity problem of the underlying classical dynamics.

\subsubsection{Regular state at fixed time}

Let us define a state defined at $t = t_s >0$, that is ``far away'' from the singularity, as follows
\begin{eqnarray}
\label{InitStateEtaV}
&&\Psi(t_s, x_1,x_2)=
\begin{cases}
0, & \text{for } x_2 \le t_H \\
v(x_1,x_2) e^{i(\Lambda_1 x_1+\Lambda_2 x_2)}, & \text{for } x_2 > t_H  ,
\end{cases}
\end{eqnarray}
where
\begin{eqnarray}
\label{InitStateEtaVExample}
&&v(x_1,x_2)=
\begin{cases}
0, & \text{for } x_2 \le t_H \\
\frac{1}{\sqrt{N_v}}
\left[\frac{x_1^2}{(\beta_1+x_1)^3} \right]
\left[\frac{(x_2-t_H)^2}{(\beta_2-t_H + x_2)^3} \right]
, & \text{for } x_2 > t_H  .
\end{cases}
\end{eqnarray}
and where $\Lambda_1, \Lambda_2, \beta_1, \beta_2 \in \dR~$ ($N_v$ denotes normalization constant).

Direct calculation gives
\begin{equation}
\label{InitStateExpPosition}
\Bra{\Psi(t_s, x_1,x_2)}\hat{q}_k\Ket{\Psi(t_s, x_1,x_2)} = \Lambda_k,~~~k = 1,2 \, ,
\end{equation}
and
\begin{equation}
\label{InitStateExpF0}
\Bra{\Psi(t_s, x_1,x_2)}\hat{F}^{(\theta)}_0 (t_s)\Ket{\Psi(t_s, x_1,x_2)} > 0 \, ,
\end{equation}
as the integrand of \eqref{InitStateExpF0} is positive definite due to  \eqref{regF2}.

One can also verify that we have
\begin{equation}
\label{InitStateExpPosition}
\Bra{\Psi(t_s, x_1,x_2)}\hat{p}_k\Ket{\Psi(t_s, x_1,x_2)} =
C_p \int_0^\infty \frac{dx_1}{x_1} \int_{t_H}^\infty dx_2
\frac{|v(x_1,x_2)|^2}{x_k(x_2-t)}< \infty ,~~~k = 1,2 \, ,
\end{equation}
where $C_p$ is a constant.

Therefore, the  state \eqref{InitStateEtaV} is regular at $t_s$.

\subsubsection{Initial state obtained in backward evolution}

The state \eqref{InitStateEtaV} is an example of the state that can be presented, due to \eqref{ns10} and \eqref{nor2},
as follows
\begin{eqnarray}
&& {\color{black}
\Psi(t_s,x_1,x_2)=0,~~\mbox{for}~~x_2 \le t_H - t_s},~~\mbox{where}~~t_s < t_H \ , \label{back1} \\
&& \Psi(t_s,x_1,x_2)=\eta(x_1,x_2+t_s) \sqrt{\frac{x_2}{x_2+t_s}}
\times  \nonumber \\
&& \times \exp\left(i \int_{0}^{t_s} K(t',x_1,x_2+t_s-t')dt' \right),~~
\mbox{for}~~ x_2 \ge t_H - t_s \label{back2} \, .
\end{eqnarray}

The above state can be inverted to get the initial state at $t = 0$  via the backward evolution:
\begin{eqnarray}
&& {\color{black}
\eta(x_1,x_2)=0,~~ \mbox{ for }~~ x_2 \le t_H} \ , \label{back3} \\
&& \eta(x_1,x_2)=  \Psi(t_s,x_1,x_2-t_s) \sqrt{\frac{x_2}{x_2 -t_s}}
\times  \nonumber \\
&& \times \exp\left(-i \int_{0}^{t_s} K(t',x_1,x_2-t')dt' \right),~~
\mbox{ for }~~ x_2 > t_H \label{back4}\ .
\end{eqnarray}
Since the ``forward'' evolution is unitary, the ``backward'' evolution is unitary as well.

\subsubsection{Regularity of the initial state}

It is easy to check that
\begin{equation}
\label{back4}
\Bra{\eta(x_1,x_2)}\hat{q}_k \Ket{\eta(x_1,x_2)} < \infty \, ,
\end{equation}
and
\begin{equation}
\label{back5}
\Bra{\eta(x_1,x_2)}\hat{p}_k \Ket{\eta(x_1,x_2)}=
C_p \int_0^\infty \frac{dx_1}{x_1} \int_{t_H}^\infty dx_2
\frac{|\Psi(t_s,x_1,x_2-t_s)|^2}{x_k(x_2-t_s)} < \infty \ ,
\end{equation}
where $k = 1,2$ and where $C_p$ is a constant. It is so because the integrands of
 \eqref{back4} and  \eqref{back5} are
positive definite functions, and due to \eqref{symqp}.

Now, let us address the issue presented by Eq. \!\eqref{qw99}. To show that this equation cannot be satisfied,
we should prove that for all $t_H >  t_s \neq 0$ we have
\begin{equation}
\label{back6}
\Bra{\eta(x_1,x_2)}\hat{F}^{(\theta)}_0 ( t_s ) \Ket{\eta(x_1,x_2)} =
\int_0^\infty \frac{dx_1}{x_1} \int_{t_H}^\infty dx_2
\frac{\breve{F}^{(\theta)}_0 (t_s)|\Psi(t_s,x_1,x_2-t_s)|^2}{x_2-t_s}> 0 \, .
\end{equation}

We exclude the case $ t_s =0$, because for $p_1 = p_2$, due to \eqref{thetka}, we have
$ F_0^{(\theta)} = 0$, which means $\hat{F}_0^{(\theta)}(0) = \hat{0}$. The ``zero operator'' cannot
represent any physical observable as its action does not lead to any physical state. Namely, it maps any state $|\psi\rangle$
into the zero vector $\hat{0}|\psi\rangle$. Such a vector cannot be normalized. Thus,  it cannot be given any probabilistic
interpretation.

Since the integrand defining Eq. \!\eqref{back6} is positive definite, the equation is satisfied, which completes the proof.

Thus, the initial state is regular, i.e., does not satisfy Eqs. \!\eqref{qw9} and \eqref{qw99}. This implies that whenever
we have a regular state far away from the singularity (which is generic case), the initial quantum state at $t = 0$ is regular so
that the quantum evolution  is well defined for any $t \geq 0$. This is a direct consequence of the unitarity of considered
quantum evolution.

\subsection{Quantum bounce}

Let us examine the issue of possible time reversal invariance of our quantum model.  In what follows, we examine the
time reversal invariance of our Schr\"{o}dinger equation and its solution.

The operator of the time reversal, $\hat{T}: \mathcal{H} \rightarrow \mathcal{H} $, is defined to be
\begin{equation}\label{qb1}
   \hat{T}\, \psi (t, x_1, x_2) = \tilde{\psi} (t, x_1, x_2) :=  \psi (- t, x_1, x_2)^\ast,~~~\mbox{where}~~~\psi \in \mathcal{H} \, ,
\end{equation}
so its complex conjugates change the sine of the time variable in a state vector.
\begin{equation}\label{qb2}
i \frac{\partial}{\partial t} \tilde{\Psi}(t,x_1,x_2) =
 \left(- i \frac{\partial }{\partial x_2}  +\frac{i}{2x_2} - K(-t,t,x_1,x_2)
 \right) \tilde{\Psi}(t,x_1,x_2)\, .
\end{equation}

The general solution to Eq. \!\eqref{qb2}, for $t < 0$,  is found to be
\begin{equation}\label{qb3}
\tilde{\Psi}(t,x_1,x_2)=\eta(x_1,x_2 + |t| - |t_0|)\, \sqrt{\frac{x_2}{x_2 + |t| - |t_0|}}
\, \exp\left(i \int_{t_0}^t K(-t',x_1,x_2-t+t')\,dt' \right) \, ,
\end{equation}
where $ |t|\geq |t_0|$, and where  $\eta (x_1, x_2):= \tilde{\Psi} (t_0, x_1,x_2)$ is the initial state.

The unitarity of the evolution (with $t_0 = 0$) leads to the condition
\begin{equation}\label{qb4}
 \eta(x_1,x_2) = 0~~~~\mbox{for}~~~~x_2 < |t_H| \, ,
\end{equation}
which corresponds to the condition \eqref{nor2}.

For $|t| < |t_H|$ we get
\begin{equation} \label{qb4}
\BraKet{\tilde{\Psi}(t)}{\tilde{\Psi}(t)} = \int_0^\infty \frac{dx_1}{x_1}  \int_{|t_H|}^\infty d{x_2}
\frac{ |\eta(x_1,x_2)|^2}{x_2} \,,
\end{equation}
which shows that the norm is time independent.

Comparing Eqs. \!\eqref{ns9} and \eqref{qb2} we can see that the dynamical
equation fails to be time reversal invariant because the Hamiltonian $\hat{H}_0$
does not have this symmetry. However, the solutions to these equations have only
different phases. Thus, the probability density is continuous at $t = 0$ (that
marks the classical singularity) due to Eqs. \!\eqref{ns10} and \eqref{qb3}
(with $|t_0| = 0$), which we call the {\it quantum bounce}.


\section{Conclusions}

Near the classical singularity the dynamics of the general Bianchi IX model
simplifies. Due to the symmetry of the physical phase space of this model, we
can apply the affine coherent states quantization method. The quantum dynamics,
described by the Schr\"{o}dinger equation, is devoid of singularities in the
sense that the expectation values of basic operators are finite during the
quantum evolution of the system. The evolution is unitary and the probability
density of our system is continuous at $t = 0$, which marks the classical
singularity.

We name the state defined by Eqs. \!\eqref{back1}--\eqref{back2} the {\it rescue
state}.  It is chosen to be regular because one expects that the quantum
state far away from the singularity is a proper quantum state.  The
Schr\"{o}dinger evolution does not lead outside the space of such states.

The choice of the fiducial state as the rescue state leads, under the
action of the affine group, to another rescue state with changed parameter
$t_H$. Namely,
\begin{equation}\label{con1}
  U(q,p)\mathcal{K}_{t_H}=\mathcal{K}_{t_H/p},~~~~p\in (0,+\infty),
\end{equation}
where $\mathcal{K}_{t_H}$ denotes the rescue space with $t_H$ parameter (for
simplicity we consider only one half plane).  In general, the fiducial
  vector is known to be a free ``parameter'' of the coherent states quantization
  formalism. However, our final conclusion does not depend on the fiducial
  vector.

The quantum Hamiltonian, $\hat{H}_0 (t,x_1,x_2)= -\hat{q}_2 - \hat{K} (t, x_1,
x_2)$, is not invariant under the affine group action as we have
\begin{eqnarray}
&& U_1(\tilde{q}_1,\tilde{p}_1)U_2(\tilde{q}_2,\tilde{p}_2)
\hat{H}_0(t,x_1,x_2)
U_1(\tilde{q}_1,\tilde{p}_1)^{-1} U_2(\tilde{q}_2,\tilde{p}_2)^{-1} \nonumber \\
&& = \hat{H}_0 (t,\tilde{p}_1 x_1,\tilde{p}_2 x_2) = -\frac{1}{\tilde{p}_2} \hat{q}_2 - \hat{K}(t,\tilde{p}_1 x_1,\tilde{p}_2 x_2) \neq
\hat{H}_0 (t,x_1,x_2) \label{con2} \, .
\end{eqnarray}
Therefore, the affine coherent states quantization does not introduce the affine
symmetry as the symmetry of our quantum system. However, the quantum evolution
restricted to the space of all possible rescue spaces is unitary and devoid of
singularities.  Thus, the rescue space, which spans the subspace of our Hilbert
space, is of basic importance in our quantization scheme.

The non-diagonal BIX underlies the BKL conjecture which concerns the generic
singularity of general relativity.  Therefore, our results suggest that {\it
quantum} general relativity is free from singularities. Classical singularity
is replaced by quantum bounce, which presents a unitary evolution of the quantum
gravity system.

Since general relativity successfully describes almost all available
gravitational data, it makes sense its quantization to get the extension to
the quantum regime. The latter could be used to describe quantum gravity
effects expected to occur near the beginning of the Universe and in the
interior of black holes.

Our fully quantum results show that the preliminary results obtained for the
diagonal BIX within the semiclassical affine coherent states approximation
\cite{Bergeron:2015lka,Bergeron:2015ppa} are correct. Appendix B presents the
affine coherent states applied in these papers, which define another
parametrization of our coherent states.

As far as we are aware, we are pioneers in addressing the issue of
resolving the {\it generic} singularity problem of general relativity via
quantization.  Therefore, trying to confirm our preliminary results within different
quantization scheme would be interesting. In particular, the choice of
different time and corresponding Hamiltonian in the reduced phase space
approach (see, Eq. \!\eqref{sim1}) would be valuable.

\acknowledgments We would like  to thank Katarzyna G\'{o}rska for the derivation of Eqs. \!\eqref{eq5}--\eqref{eq7},
Vladimir Belinski and John Klauder for helpful correspondence, and Claus Kiefer for discussion.

\appendix

\section{Alternative affine coherent states for half-plane}
\label{alternative}

The phase space $\Pi_1$ may be identified with the affine group
$\Group{Aff}(\dR)$ by defining the multiplication law as follows
\begin{equation}\label{c1}
 (q^\prime, p^\prime)\cdot (q, p) =(\frac{q}{p^\prime}+ q^\prime, p^\prime p ),
\end{equation}
with the unity $(0,1)$ and the inverse
\begin{equation}\label{c2}
(q^\prime, p^\prime)^{-1} = (-q^\prime p^\prime, \frac{1}{p^\prime}).
\end{equation}
The affine group has two, nontrivial, inequivalent irreducible unitary
representations \cite{Gel} and \cite{AK1,AK2}.  Both are realized in the Hilbert space
$L^2(\dR_+, d\nu(x))$, where $d\nu(x)=dx/x$ is the invariant measure on
the multiplicative group $(\dR_+,\cdot)$. In what follows we choose the one
defined by
\begin{equation}\label{im1}
 U(q,p)\psi(x)= e^{i q x} \psi(x/p)\, ,
\end{equation}
where $|\psi\rangle \in L^2(\dR_+, d\nu(x))$.

For simplicity of notation, let us define integrals over the affine group
$\Group{Aff}(\dR)$ as follows:
\begin{eqnarray}
&&\int_{\Group{ Aff}(\dR)} d\mu_L(q,p)= \frac{1}{2\pi}
\int_{-\infty}^{+\infty} dq \int_{0}^\infty \frac{dp}{p^2} \, ,\\
&&\int_{\Group{Aff}(\dR)} d\mu_R(q,p)= \frac{1}{2\pi}
\int_{-\infty}^{+\infty} dq \int_{0}^\infty \frac{dp}{p} \, ,\\
&&\int_{\Group{Aff}(\dR)} d\mu_U(q,p)= \frac{1}{2\pi}
\int_{-\infty}^{+\infty} dq \int_{0}^\infty dp \, \rho(q,p) \, .
\end{eqnarray}
The last one is intended to be used as invariant measure in respect
with the action $U(q,p)$.

Fixing the normalized vector $\Ket{\Phi} \in L^2(\dR_+, d\nu(x))$, called the
{\it fiducial} vector, one can define a continuous family of {\it affine}
coherent states $\Ket{q,p} \in L^2(\dR_+, d\nu(x))$ as follows
\begin{equation}\label{im22}
\Ket{q,p} = U(q,p) \Ket{\Phi}.
\end{equation}
As we have three  measures,  one can define three operators which
potentially can leads to the unity in the space $L^2(\dR_+, d\nu(x))$:
\begin{eqnarray}\label{unityOneTwo}
&& B_L=\int_{\Group{ Aff}(\dR)} d\mu_L(q,p) \Ket{q,p}\Bra{q,p} \, ,\\
&& B_R=\int_{\Group{ Aff}(\dR)} d\mu_R(q,p) \Ket{q,p}\Bra{q,p} \, ,\\
&& B_U= \int_{\Group{Aff}(\dR)} d\mu_U(q,p) \, \rho(q,p) \Ket{q,p}\Bra{q,p} \, .
\end{eqnarray}
Let us check which one is invariant under the action $U(q,p)$ of the affine
group:
\begin{equation} \label{BLInvU}
U(q',p') B_U U(q',p')^\dagger =
\int_{-\infty}^{+\infty} dq \int_0^\infty dp \rho(q,p)
\Ket{q/p'+q',p'p}\Bra{q/p'+q',p'p}
\end{equation}
One needs to replace the variables under the integral:
\begin{eqnarray} \label{BLInvU2}
&&\tilde{q}=q/p'+q' \quad \mbox{and} \quad \tilde{p}=p'p \, ,\\
&& q=p'(\tilde{q}-q') \quad \mbox{and} \quad p=\frac{\tilde{p}}{p'}\, .
\end{eqnarray}
Calculating the Jacobian
$\frac{\partial(q,p)}{\partial(\tilde{q},\tilde{p})}=1$ one gets:
\begin{equation} \label{BLInvU3}
d\mu_U(q,p)=\rho(q,p)  d\tilde{q}d\,\tilde{p}=
\rho(p'(\tilde{q}-q'),\frac{\tilde{p}}{p'}) d\tilde{q}d\,\tilde{p}
\end{equation}
This implies, the transformed weight should be equal to the initial one,
$\rho(p'(\tilde{q}-q'),\frac{\tilde{p}}{p'})=\rho(q,p)$ for every
$(q',p')$. The simplest solution is $\rho(q,p)=\mathrm{const}$, so we get
$d{\mu_U}(q,p)=dq\,dp$.

It also implies that the operators $B_L$ and $B_R $ do not commute with the
affine group. The action \eqref{im1} is not compatible neither with the left
invariant, nor with right invariant measures on the affine group.

The {\it irreducibility} of the representation, used to define the coherent
states \eqref{im22}, enables making use of Schur's lemma \cite{BR}, which leads
to the resolution of the unity in $L^2(\dR_+, d\nu(x))$:
\begin{equation}\label{im44}
\int_{\Group{ Aff}(\dR)}  d{\mu_U}(q,p) \Ket{q,p}\Bra{q,p} = A_\Phi\;\I \; ,
\end{equation}
where the constant $A_\Phi$ can be calculated using any arbitrary, normalized
vector $\Ket{f} \in L^2(\dR_+, d\nu(x))$:
\begin{equation}\label{im3b}
A_\Phi = \int_{\Group{ Aff}(\dR)}d{\mu_U}(q,p)\,
\BraKet{f}{q,p}\BraKet{q,p}{f} \, .
\end{equation}
This formula can be calculated directly:
\begin{eqnarray}\label{im3b2}
&& A_\Phi = \int_{\Group{ Aff}(\dR)}  d{\mu_U}(q,p) \nonumber \\
&&\times  \int_0^\infty d\nu(x')\int_0^\infty d\nu(x)
(f(x')^\star e^{iqx'}\Phi(x'/p))(e^{-iqx}\Phi(x/p)^\star f(x)) \nonumber \\
&& = \int_0^\infty \frac{dx'}{x'}\int_0^\infty \frac{dx}{x}
\int_0^\infty dp
\left[\frac{1}{2\pi}\int_{-\infty}^{+\infty} dq e^{iq(x'-x)}\right]
f(x')^\star f(x) \Phi(x'/p)\Phi(x/p)^\star \nonumber \\
&& = \int_0^\infty \frac{dx}{x^2} |f(x)|^2
\int_0^\infty dp |\Phi(x/p)|^2    \nonumber \\
&& = \left(\int_0^\infty \frac{dx}{x} |f(x)|^2\right)
\left(\int_0^\infty \frac{dp}{p^2} |\Phi(p)|^2 \right)
=\int_0^\infty \frac{dp}{p^2} |\Phi(p)|^2 \,
\end{eqnarray}
if $\BraKet{f}{f}=1$.

\noindent In the derivation of \eqref{im3b2} we have used the equations:
\begin{equation}\label{rem}
  \langle x | x^\prime\rangle =
x \delta (x - x^\prime),~~~\int_0^\infty \frac{dx}{x}\,|x \rangle \langle x |
= \id,
~~~\int_0^\infty \frac{dx}{x}\,\delta (x - x^\prime)f(x) = f(x^\prime) \, .
\end{equation}

\section{Orthonormal basis of the carrier space} \label{basis}

The  basis of the Hilbert space $L^2(\dR_+, d\nu(x))$ is known  to be \cite{GM}
\begin{equation}\label{aa1}
e^{(\alpha)}_n (x) =
\sqrt{\frac{n!}{\Gamma (n + \alpha + 1)}}\,e^{-x/2} x^{(1 +\alpha)/2}\,L_n^{(\alpha)}(x),
\end{equation}
where $L_n^{(\alpha)}$ is the Laguerre function and  $\alpha > -1$.  One can verify that $\int_0^\infty
e^{(\alpha)}_n (x) e^{(\alpha)}_m (x) d\nu(x)= \delta_{n m}$ so that
$e^{(\alpha)}_n (x)$ is an orthonormal basis and can be used in calculations.


\end{document}